\documentclass{article}
\usepackage{graphicx} 
\usepackage[colorlinks]{hyperref}
\hypersetup{colorlinks=true,
            citecolor=blue,
            urlcolor=blue,}
\usepackage{amsmath}
\usepackage{amsfonts}
\usepackage{authblk}
\usepackage[margin=1.2in]{geometry}
\usepackage{tikz}
\usetikzlibrary{shapes.geometric}
\usetikzlibrary{decorations.pathreplacing}
\usepackage{algpseudocode}
\usepackage{algorithm}
\usepackage{xcolor}
\usepackage{colortbl}
\usepackage{nicematrix}
\usepackage{censor}

\usepackage{soul} 

\usepackage{pgfplots}
\pgfplotsset{compat=1.18}

\title{Faster Polyhedral Intersections in 3D}
\author[1,3]{Joseph Donato \thanks{jdonato@lanl.gov}}
\author[2]{Daniel Shevitz \thanks{danny.shevitz@yahoo.com}}
\author[1]{Rao Garimella \thanks{rao@lanl.gov}}
\affil[1]{Los Alamos National Laboratory}
\affil[2]{Los Alamos National Laboratory (Ret.)}
\affil[3]{Dartmouth College}
\date{}

\begin{document}
\maketitle

\begin{abstract}
The operation of intersecting two polyhedra, which can be reduced to clipping one polyhedron with a series of planes formed from another polyhedron, is central to intersection based mesh interpolation and material interface reconstruction.  R3D is a robust library for such operation which implements the algorithm of Sugihara.  In this paper, we present J3D, a new implementation of the clipping algorithm of Sugihara that provides substantial performance improvements over R3D.
\end{abstract}
\textbf{Keywords:} Polyhedral clipping, Computational physics, Algorithms, Data structures

\section{Introduction}
In the implementation of fluid simulations, especially when shocks are involved, the Arbitrary Lagrangian Eulerian (ALE) method is heavily relied upon to reliably simulate flows \cite{BARLOW2016603,LOUBERE20104724}.
At a high level, the ALE method operates as follows.
First, the mesh is deformed with the fluid until it is no longer viable.
Then, to continue the simulation, the fields of the deformed mesh are conservatively interpolated onto a higher quality mesh and the simulation continues.
A commonly used method to conservatively interpolate the field values from a source mesh to a target mesh is intersection based remap \cite{Herring-2021}.
In this procedure, each cell of the target mesh is intersected with cells of the source mesh, and the resulting moments of intersection are used as weights in the interpolation of source cell values onto the target cell.
Intersecting two polyedra reduces to the problem of assuming that one of them is convex and intersecting the planes of the convex polyhedron with the other. 
Another operation commonly operation found in ALE simulations is material interface reconstruction \cite{VOF,dyadechko2005moment,LVIRA,lviraplus}.
In interface reconstruction, the volume fractions and, possibly centroids, of materials in a cell must be used to recover the geometry of the interfaces between the materials.
This is done by repeatedly clipping the polyhedron with a plane to find the optimal material geometry.
As suggested from the descriptions of the two aforementioned procedures, intersecting a polyhedron with a plane is fundamental.
In fact, while polyhedral intersection is typically a very small algorithm within the context of a larger ALE method, it has significant performance impacts.
This is exacerbated by the fact that as the mesh deforms in ALE, the faces of a polyhedron may no longer be planar.
To resolve this, faces of a polyehedron are ``facetized" which adds to the problem complexity.
That is, the faces of one polyhedron are decomposed into triangular faces by adding a point in the center of each face and joining it with each of the edges making up the original face.

\begin{figure}[H]
\begin{center}
\begin{minipage}{.3\textwidth}
\begin{tikzpicture}
    \node[circle,fill,inner sep=1pt] (1) at (0,0) {};
    \node[circle,fill,inner sep=1pt] (2) at (2,0) {};
    \node[circle,fill,inner sep=1pt] (3) at (0,2) {};
    \node[circle,fill,inner sep=1pt] (4) at (2,2) {};
    \node[circle,fill,inner sep=1pt] (5) at (1.3,1) {};
    \node[circle,fill,inner sep=1pt] (6) at (3.3,1) {};
    \node[circle,fill,inner sep=1pt] (7) at (1.3,3) {};
    \node[circle,fill,inner sep=1pt] (8) at (3.3,3) {};
    
    \path [] (1) edge node[] {} (2); 
    \path [] (1) edge node[] {} (3);
    \path [] (3) edge node[] {} (4);
    \path [] (2) edge node[] {} (4);
    
    \path [dashed] (5) edge node[] {} (6); 
    \path [dashed] (5) edge node[] {} (7);
    \path [] (7) edge node[] {} (8);
    \path [] (6) edge node[] {} (8);
    
    \path [dashed] (1) edge node[] {} (5); 
    \path [] (2) edge node[] {} (6);
    \path [] (3) edge node[] {} (7);
    \path [] (4) edge node[] {} (8);
\end{tikzpicture}
\end{minipage}
\begin{minipage}{.3\textwidth}
\begin{tikzpicture}
    \node[circle,fill,inner sep=1pt] (1) at (0,0) {};
    \node[circle,fill,inner sep=1pt] (2) at (2,0) {};
    \node[circle,fill,inner sep=1pt] (3) at (0,2) {};
    \node[circle,fill,inner sep=1pt] (4) at (2,2) {};
    \node[circle,fill,inner sep=1pt] (5) at (1.3,1) {};
    \node[circle,fill,inner sep=1pt] (6) at (3.3,1) {};
    \node[circle,fill,inner sep=1pt] (7) at (1.3,3) {};
    \node[circle,fill,inner sep=1pt] (8) at (3.3,3) {};
    \node[circle,fill,inner sep=1pt] (9) at (1,1) {};
    \node[circle,fill,inner sep=1pt] (10) at (2.3,2) {};
    \node[circle,fill,inner sep=1pt] (11) at (2.65,1.5) {};
    \node[circle,fill,inner sep=1pt] (12) at (0.65,1.5) {};
    \node[circle,fill,inner sep=1pt] (13) at (1.65,2.5) {};
    \node[circle,fill,inner sep=1pt] (14) at (1.65,0.5) {};
    \path [] (1) edge node[] {} (2); 
    \path [] (1) edge node[] {} (3);
    \path [] (3) edge node[] {} (4);
    \path [] (2) edge node[] {} (4);
    \path [] (1) edge node[] {} (9);
    \path [] (2) edge node[] {} (9);
    \path [] (3) edge node[] {} (9);
    \path [] (4) edge node[] {} (9);
    \path [dashed] (5) edge node[] {} (6); 
    \path [dashed] (5) edge node[] {} (7);
    \path [] (7) edge node[] {} (8);
    \path [] (6) edge node[] {} (8);
    \path [dashed] (5) edge node[] {} (10);
    \path [dashed] (6) edge node[] {} (10);
    \path [dashed] (7) edge node[] {} (10);
    \path [dashed] (8) edge node[] {} (10);
    \path [] (2) edge node[] {} (11);
    \path [] (4) edge node[] {} (11);
    \path [] (6) edge node[] {} (11);
    \path [] (8) edge node[] {} (11);
    \path [dashed] (1) edge node[] {} (12);
    \path [dashed] (3) edge node[] {} (12);
    \path [dashed] (5) edge node[] {} (12);
    \path [dashed] (7) edge node[] {} (12);
    \path [] (3) edge node[] {} (13);
    \path [] (4) edge node[] {} (13);
    \path [] (7) edge node[] {} (13);
    \path [] (8) edge node[] {} (13);
    \path [dashed] (1) edge node[] {} (14);
    \path [dashed] (2) edge node[] {} (14);
    \path [dashed] (5) edge node[] {} (14);
    \path [dashed] (6) edge node[] {} (14);
    \path [dashed] (1) edge node[] {} (5); 
    \path [] (2) edge node[] {} (6);
    \path [] (3) edge node[] {} (7);
    \path [] (4) edge node[] {} (8);
\end{tikzpicture}
\end{minipage}
\caption{A hexahedra (left) and the same hexahedra after being facetized into triangular faces (right).}
\end{center}
\end{figure}
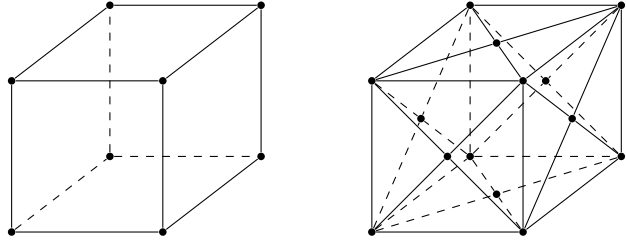

Now, given the importance of clipping a polyhedron efficiently, the focus of this paper will be just that.
We will begin by briefly describing existing algorithms which do this and why we chose the algorithm of Sugihara \cite{Sugihara} in  particular.
We will then go into detail of an existing implementation known as R3D \cite{powell2015r3d} as well as its limitations.
Finally, we will describe our novel implementation known as J3D, the data structures we introduce for it, and the significant performance improvements it yields.
The code repository of J3D is located at \url{https://github.com/lanl/j3d}.

\section{Background}
The landscape of algorithms to intersect two polyhedra is already rich.
Existing algorithms offer varying results in terms of asymptotic performance, floating point accuracy, and practicality.
Algorithms such as \cite{MULLER1978217,Chazelle} while providing impressive asymptotic performance, are difficult to implement and are thus not considered for practical applications.

On the other hand, algorithms that use a Binary Space Partitioning (BSP) Tree like \cite{Bernstein} are numerically accurate and fast for very large geometries.  However, they require the initialization of a BSP-tree which is an overhead we cant afford since our use cases require intersections to be called many times on polyhedra known to be relatively small.

Ultimately, we settled on Sugihara's clipping algorithm \cite{Sugihara} which we also refer to as clip-n-cap.  The algorithm is very simple and practical to implement.  Moreover, to address robustness, the only floating point operations required are to determine if a vertex is on the positive or negative side of the plane.  It is well known that differences in classification of a vertex as being in front or behind the plane in the Sugihara algorithm can result in a different topology for the polyhedron. In particular, when clipping non-convex polyhedra, these errors can yield a resulting graph with small independent polyhedral pieces with possibly negative volumes. If the only goal is to extract out the moments of intersection as needed in conservative remapping, these small positive or negative poyhedral fragments do not make a significant difference.  On the other hand, in the case of interface reconstruction, if one needs to recover the polyhedron topology for further reuse, then one needs to do some cleanup of the small fragments using a tolerance which only needs to be done once after clipping with all the planes is done.

In the Sugihara clipping algorithm \cite{Sugihara}, the polyhedron is represented as a graph $G=(V,E)$ where $V$ is the set of vertices and $E$ is the set of edges.  When a plane ``clips" $G$, the vertices are partitioned into two sets $V_{1}$ and $V_{2}$.  $V_{1}$ is the set of vertices on the positive side of the plane and $V_{2}$ is the set of points on the negative side of the plane.  We reconstruct the polyhedron on the positive side of the plane as follows.
\begin{enumerate}
    \item Each edge with one endpoint in $V_{1}$ and one endpoint in $V_{2}$ is split in two by a new vertex generated at the intersection of the edge and the clipping plane.  
    \item Next, a cycle separating $V_{1}$ and $V_{2}$ is generated by connecting the aforementioned new vertices.  This cycle represents the capping face.  
    \item Finally, the vertices in $V_{2}$ and their incident edges are removed from the graph.
\end{enumerate}

With the Sugihara algorithm in mind, we will now introduce two common representations of a polyhedron.  A boundary representation (BREP) describes a hierarchy between a cell, its faces, and the vertices on each face ordered in the outward normal direction to the face.  The graph representation of a polyhedron is one in which each vertex maintains a list of edge connected nodes also referred to as neighbors.

\begin{table}[H]
\centering
\begin{tikzpicture}[baseline=(current bounding box.center)]
    \node[draw, circle, inner sep=1pt,minimum size=0.15cm] (1) at (0,0) {1};
    \node[draw, circle, inner sep=1pt,minimum size=0.15cm] (2) at (2,0) {2};
    \node[draw, circle, inner sep=1pt,minimum size=0.15cm] (3) at (0,2) {3};
    \node[draw, circle, inner sep=1pt,minimum size=0.15cm] (4) at (2,2) {4};
    \node[draw, circle, inner sep=1pt,minimum size=0.15cm] (5) at (1,1) {5};
    \node[draw, circle, inner sep=1pt,minimum size=0.15cm] (6) at (3,1) {6};
    \node[draw, circle, inner sep=1pt,minimum size=0.15cm] (7) at (1,3) {7};
    \node[draw, circle, inner sep=1pt,minimum size=0.15cm] (8) at (3,3) {8};
    \node[] (9) at (1.5,1.5) {f1};
    \node[] (10) at (2.5,1.9) {f2};
    \node[] (11) at (-0.5,1.6) {f3};
    \node[] (12) at (3.5,1.9) {f4};
    \node[] (12) at (1.5,3.5) {f5};
    \node[] (13) at (1.5,-0.5) {f6};
    \path [] (1) edge node[] {} (2); 
    \path [] (1) edge node[] {} (3);
    \path [] (3) edge node[] {} (4);
    \path [] (2) edge node[] {} (4);
    \path [dashed] (5) edge node[] {} (6); 
    \path [dashed] (5) edge node[] {} (7);
    \path [] (7) edge node[] {} (8);
    \path [] (6) edge node[] {} (8);
    \path [dashed] (1) edge node[] {} (5); 
    \path [] (2) edge node[] {} (6);
    \path [] (3) edge node[] {} (7);
    \path [] (4) edge node[] {} (8);
\end{tikzpicture}
\hspace{0.5cm}
\begin{tabular}{@{}cc@{}}
\hline
\multicolumn{2}{c}{BREP}\\ 
\hline
face & vertices   \\
f1     & [1,2,4,3]  \\
f2     & [7,8,6,5]  \\
f3     & [3,7,5,1]  \\
f4     & [8,4,2,6]  \\
f5     & [3,4,8,7]  \\
f6     & [1,5,6,2]  \\
\hline
\end{tabular}
\begin{tabular}{@{}cc@{}}
\hline
\multicolumn{2}{c}{Graph}\\ 
\hline
vertex & neighbors   \\
1     & [3,5,2] \\
2     & [4,1,6] \\
3     & [7,1,4] \\
4     & [8,3,2] \\
5     & [7,6,1] \\
6     & [8,2,5] \\
7     & [8,5,3] \\
8     & [4,6,7] \\
\hline
\end{tabular}
\caption{A hexahedron with labeled vertices and faces.  The left table demonstrates the corresponding BREP with faces and their nodes ordered in the outward normal direction.  The right table demonstrates the corresponding graph with nodes and their neighbors ordered counter-clockwise relative to the exterior of the the hexahedron.}
\end{table}

R3D and J3D both represent the polyhedron in the graph representation with the additional restriction that neighbors around a node are ordered counter-clockwise relative to the exterior of the polyhedron.  The ordering is required so that clipped faces after step $1$ are traversed correctly and to compute the moments of a polyhedron.     

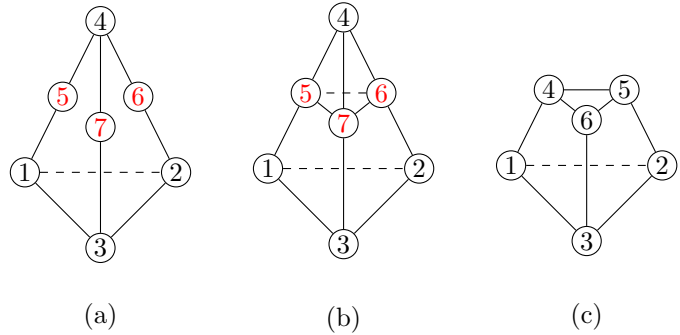
\begin{figure}[H]
\begin{center}
\begin{minipage}{.2\textwidth}
\begin{tikzpicture}[scale=1.0]
    \node[] (3) at (1,-1.9) {(a)};
    \node[draw, circle, inner sep=1pt,minimum size=0.15cm] (1) at (0,0) {1};
    \node[draw, circle, inner sep=1pt,minimum size=0.15cm] (2) at (2.0,0) {2};
    \node[draw, circle, inner sep=1pt,minimum size=0.15cm] (3) at (1,-1) {3};
    \node[draw, circle, inner sep=1pt,minimum size=0.15cm] (4) at (1,2) {4};
    \node[draw, circle, inner sep=1pt,minimum size=0.15cm] (5) at (0.5,1.0) {\textcolor{red}{5}};
    \node[draw, circle, inner sep=1pt,minimum size=0.15cm] (6) at (1.5,1.0) {\textcolor{red}{6}};
    \node[draw, circle, inner sep=1pt,minimum size=0.15cm] (7) at (1,0.6) {\textcolor{red}{7}};
    
    \path [dashed] (1) edge node[] {} (2); 
    \path [] (1) edge node[] {} (3);
    \path [] (2) edge node[] {} (3);
    
    \path [] (1) edge node[] {} (5);
    \path [] (2) edge node[] {} (6);
    \path [] (3) edge node[] {} (7);

    \path [] (5) edge node[] {} (4);
    \path [] (6) edge node[] {} (4);
    \path [] (7) edge node[] {} (4); 
\end{tikzpicture}
\end{minipage}
\begin{minipage}{.2\textwidth}
\begin{tikzpicture}[scale=1.0]
    \node[] (3) at (1,-2) {(b)};
    \node[draw, circle, inner sep=1pt,minimum size=0.15cm] (1) at (0,0) {1};
    \node[draw, circle, inner sep=1pt,minimum size=0.15cm] (2) at (2,0) {2};
    \node[draw, circle, inner sep=1pt,minimum size=0.15cm] (3) at (1,-1) {3};
    \node[draw, circle, inner sep=1pt,minimum size=0.15cm] (4) at (1,2) {4};
    \node[draw, circle, inner sep=1pt,minimum size=0.15cm] (5) at (0.5,1.0) {\textcolor{red}{5}};
    \node[draw, circle, inner sep=1pt,minimum size=0.15cm] (6) at (1.5,1.0) {\textcolor{red}{6}};
    \node[draw, circle, inner sep=1pt,minimum size=0.15cm] (7) at (1,0.6) {\textcolor{red}{7}};
    
    \path [dashed] (1) edge node[] {} (2); 
    \path [] (1) edge node[] {} (3);
    \path [] (2) edge node[] {} (3);
    
    \path [] (1) edge node[] {} (5);
    \path [] (2) edge node[] {} (6);
    \path [] (3) edge node[] {} (7);

    \path [] (5) edge node[] {} (7);
    \path [] (7) edge node[] {} (6);
    \path [dashed] (6) edge node[] {} (5);

    \path [] (5) edge node[] {} (4);
    \path [] (7) edge node[] {} (4);
    \path [] (6) edge node[] {} (4);
\end{tikzpicture}
\end{minipage}
\begin{minipage}{.2\textwidth}
\begin{tikzpicture}[scale=1.0]
    \node[] (3) at (1,-2) {(c)};
    \node[draw, circle, inner sep=1pt,minimum size=0.15cm] (1) at (0,0) {1};
    \node[draw, circle, inner sep=1pt,minimum size=0.15cm] (2) at (2,0) {2};
    \node[draw, circle, inner sep=1pt,minimum size=0.15cm] (3) at (1,-1) {3};
    \node[] (7) at (1,2) {};
    \node[draw, circle, inner sep=1pt,minimum size=0.15cm] (4) at (0.5,1.0) {4};
    \node[draw, circle, inner sep=1pt,minimum size=0.15cm] (5) at (1.5,1.0) {5};
    \node[draw, circle, inner sep=1pt,minimum size=0.15cm] (6) at (1,0.6) {6};
    
    \path [dashed] (1) edge node[] {} (2); 
    \path [] (1) edge node[] {} (3);
    \path [] (2) edge node[] {} (3);
    
    \path [] (1) edge node[] {} (4);
    \path [] (2) edge node[] {} (5);
    \path [] (3) edge node[] {} (6);

    \path [] (4) edge node[] {} (5);
    \path [] (5) edge node[] {} (6);
    \path [] (4) edge node[] {} (6);
\end{tikzpicture}
\end{minipage}
\end{center}
\caption{The clip-n-cap algorithm divided into three stages.  (a) New vertices (in red) are generated on each edge connecting a vertex from $V_{1}=\{1,2,3\}$ to a vertex from $V_{2}=\{4\}$.  (b) A cycle $(5,6,7)$ separating $V_{1}$ and $V_{2}$ is generated by connecting the new vertices.  (c) The vertices in $V_{2}$ and their incident edges are removed from the graph and vertices are relabeled.}
\end{figure}

Meshes typically represent polyhedra in the BREP and are converted into the graph representation for clip-n-cap.  This is because only local information (neighbors) of vertices have to be modified during clip-n-cap in the graph representation.  To compute the moments, for each face, one begins with picking a starting vertex $A$ on that face.  Then, for each directed edge $(B,C)$ on the face not containing $A$, the points $A$, $B$, and $C$ are use to construct a triangular facet.  Finally, each tuple of three points forming a triangular facet along with the origin are used to build a tetrahedron to be integrated over \cite{6127882}.  This is very straightforward and performant in the BREP because BREPs inherently have the notion of a faces described by a sequence of vertices from which we can pick out pairs of consecutive vertices. On the other hand, a graph representation does not have such a natural face representation.

\begin{figure}[H]
\begin{center}
\begin{minipage}{.3\textwidth}
\begin{tikzpicture}
    \node[] (7) at (0.5,-1) {(a)};

    \node[draw, circle, inner sep=1pt,minimum size=0.15cm] (1) at (0,0) {1};
    \node[draw, circle, inner sep=1pt,minimum size=0.15cm] (2) at (1.5,0) {2};
    \node[draw, circle, inner sep=1pt,minimum size=0.15cm] (5) at (-0.4,1.4) {5};
    \node[draw, circle, inner sep=1pt,minimum size=0.15cm] (3) at (1.9,1.4) {3};
    \node[draw, circle, inner sep=1pt,minimum size=0.15cm] (4) at (0.75,2.3) {4};

    \node[draw, circle, inner sep=1pt,minimum size=0.15cm] (6) at (-1.0, 0.3) {$O$};

    \path [] (1) edge node[] {} (2);
    \path [] (2) edge node[] {} (3);
    \path [] (3) edge node[] {} (4);
    \path [] (4) edge node[] {} (5);
    \path [] (5) edge node[] {} (1);

    \path [dashed] (1) edge node[] {} (3);
    \path [dashed] (1) edge node[] {} (4);

    \path [red] (6) edge node[] {} (1);
    \path [red] (6) edge node[] {} (2);
    \path [red] (6) edge node[] {} (3);
    
    \draw[ultra thick, ->] (0.75,0.2) arc (-80:190:0.7);
\end{tikzpicture}
\end{minipage}
\begin{minipage}{.3\textwidth}
\begin{tikzpicture}
    \node[] (7) at (0.5,-1) {(b)};

    \node[draw, circle, inner sep=1pt,minimum size=0.15cm] (1) at (0,0) {1};
    \node[draw, circle, inner sep=1pt,minimum size=0.15cm] (2) at (1.5,0) {2};
    \node[draw, circle, inner sep=1pt,minimum size=0.15cm] (5) at (-0.4,1.4) {5};
    \node[draw, circle, inner sep=1pt,minimum size=0.15cm] (3) at (1.9,1.4) {3};
    \node[draw, circle, inner sep=1pt,minimum size=0.15cm] (4) at (0.75,2.3) {4};

    \node[draw, circle, inner sep=1pt,minimum size=0.15cm] (6) at (-1.0, 0.3) {$O$};

    \path [] (1) edge node[] {} (2);
    \path [] (2) edge node[] {} (3);
    \path [] (3) edge node[] {} (4);
    \path [] (4) edge node[] {} (5);
    \path [] (5) edge node[] {} (1);

    \path [dashed] (1) edge node[] {} (3);
    \path [dashed] (1) edge node[] {} (4);

    \path [red] (6) edge node[] {} (1);
    \path [red] (6) edge node[] {} (3);
    \path [red] (6) edge node[] {} (4);
    
    \draw[ultra thick, ->] (0.75,0.2) arc (-80:190:0.7);
\end{tikzpicture}
\end{minipage}
\begin{minipage}{.3\textwidth}
\begin{tikzpicture}
    \node[] (7) at (0.5,-1) {(c)};

    \node[draw, circle, inner sep=1pt,minimum size=0.15cm] (1) at (0,0) {1};
    \node[draw, circle, inner sep=1pt,minimum size=0.15cm] (2) at (1.5,0) {2};
    \node[draw, circle, inner sep=1pt,minimum size=0.15cm] (5) at (-0.4,1.4) {5};
    \node[draw, circle, inner sep=1pt,minimum size=0.15cm] (3) at (1.9,1.4) {3};
    \node[draw, circle, inner sep=1pt,minimum size=0.15cm] (4) at (0.75,2.3) {4};

    \node[draw, circle, inner sep=1pt,minimum size=0.15cm] (6) at (-1.0, 0.3) {$O$};

    \path [] (1) edge node[] {} (2);
    \path [] (2) edge node[] {} (3);
    \path [] (3) edge node[] {} (4);
    \path [] (4) edge node[] {} (5);
    \path [] (5) edge node[] {} (1);

    \path [dashed] (1) edge node[] {} (3);
    \path [dashed] (1) edge node[] {} (4);

    \path [red] (6) edge node[] {} (1);
    \path [red] (6) edge node[] {} (4);
    \path [red] (6) edge node[] {} (5);
    
    \draw[ultra thick, ->] (0.75,0.2) arc (-80:190:0.7);
\end{tikzpicture}
\end{minipage}
\end{center}
\caption{The procedures (a), (b), then (c) of constructing tetrahedra from the pentagonal face of a polyhedron to compute its volume.  A starting vertex 1, is picked.  As we walk around the face, the segments (2,3), (3,4), and (4,5) are used along with vertex 1 to build triangular faces.  Each of these triangular faces along with the origin $O$ are used to build tetrahedra and their volumes are summed to compute the volume of the polyhedron.}
\end{figure}
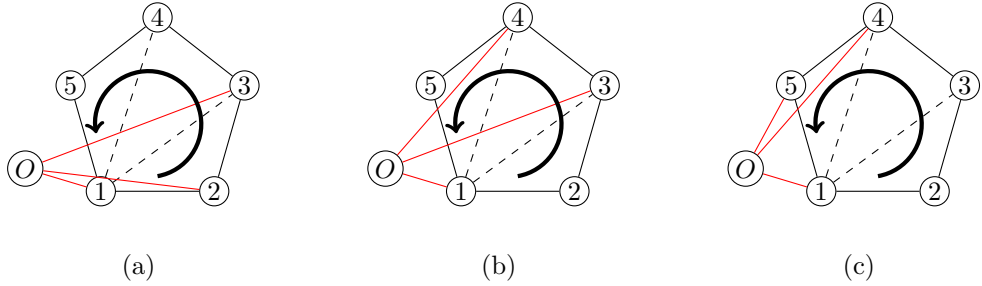

\section{R3D}
R3D is a C library which was introduced to conduct conservative voxelization and later refined at Los Alamos National Laboratory for more general polyhedral clipping \cite{POWELL2015340, powell2015r3d}.  Here, we describe some of the details of the R3D implementation of clip-n-cap and their effect on performance.  R3D represents its polyhedra as a graph but imposes an additional restriction that every node has a valence of three, which is not present in the original Sugihara paper.  This restriction was introduced to enhance vectorization of computations but handicaps the algorithm in other ways. Most notably, if a polyhedron has vertices with a valence greater than three, the algorithm has to duplicate vertices and create edges to initialize the internal representation (Figure 4).

\begin{figure}[H]
\begin{center}
\begin{minipage}{.2\textwidth}
\begin{tikzpicture}[scale=0.7]
    \node[] (10) at (0,-2.5) {(a)};
    \node[draw, circle, inner sep=1pt,minimum size=0.15cm] (1) at (0,0) {1};
    \node[draw, circle, inner sep=1pt,minimum size=0.15cm] (2) at (1.5,1.5) {2};
    \node[draw, circle, inner sep=1pt,minimum size=0.15cm] (3) at (-1.5,1.5) {3};
    \node[draw, circle, inner sep=1pt,minimum size=0.15cm] (4) at (-1.5,-1.5) {4};
    \node[draw, circle, inner sep=1pt,minimum size=0.15cm] (5) at (1.5,-1.5) {5};
    \path [] (1) edge node[] {} (2);
    \path [] (1) edge node[] {} (3);
    \path [] (1) edge node[] {} (4);
    \path [] (1) edge node[] {} (5); 
\end{tikzpicture}
\end{minipage}
\begin{minipage}{.2\textwidth}
\begin{tikzpicture}[scale=0.7]
    \node[] (10) at (0,-2.5) {(b)};
    \node[draw, circle, inner sep=1pt,minimum size=0.15cm] (2) at (1.5,1.5) {2};
    \node[draw, circle, inner sep=1pt,minimum size=0.15cm] (3) at (-1.5,1.5) {3};
    \node[draw, circle, inner sep=1pt,minimum size=0.15cm] (4) at (-1.5,-1.5) {4};
    \node[draw, circle, inner sep=1pt,minimum size=0.15cm] (5) at (1.5,-1.5) {5};
    \node[draw, circle, inner sep=1pt,minimum size=0.15cm] (6) at (0.75,0.75) {6};
    \node[draw, circle, inner sep=1pt,minimum size=0.15cm] (7) at (-0.75,0.75) {7};
    \node[draw, circle, inner sep=1pt,minimum size=0.15cm] (8) at (-0.75,-0.75) {8};
    \node[draw, circle, inner sep=1pt,minimum size=0.15cm] (9) at (0.75,-0.75) {9};
    \path [] (3) edge node[] {} (7);
    \path [] (2) edge node[] {} (6);
    \path [] (5) edge node[] {} (9);
    \path [] (4) edge node[] {} (8);
    \path [] (7) edge node[] {} (6);
    \path [] (7) edge node[] {} (8);
    \path [] (6) edge node[] {} (9);
    \path [] (8) edge node[] {} (9);
\end{tikzpicture}
\end{minipage}
\begin{minipage}{.2\textwidth}
\begin{tikzpicture}[scale=0.7]
    \node[] (10) at (0,-2.5) {(c)};
    \node[draw, circle, inner sep=1pt,minimum size=0.15cm] (2) at (1.5,1.5) {2};
    \node[draw, circle, inner sep=1pt,minimum size=0.15cm] (3) at (-1.5,1.5) {3};
    \node[draw, circle, inner sep=1pt,minimum size=0.15cm] (4) at (-1.5,-1.5) {4};
    \node[draw, circle, inner sep=1pt,minimum size=0.15cm] (5) at (1.5,-1.5) {5};
    \node[draw, circle, inner sep=1pt,minimum size=0.15cm] (6) at (0.75,0.75) {6};
    \node[draw, circle, inner sep=1pt,minimum size=0.15cm] (7) at (-0.75,0.75) {7};
    \path [] (3) edge node[] {} (7);
    \path [] (2) edge node[] {} (6);
    \path [] (7) edge node[] {} (6);
\end{tikzpicture}
\end{minipage}
\begin{minipage}{.2\textwidth}
\begin{tikzpicture}[scale=0.7]
    \node[] (10) at (0,-2.5) {(d)};
    \node[draw, circle, inner sep=1pt,minimum size=0.15cm] (2) at (1.5,1.5) {2};
    \node[draw, circle, inner sep=1pt,minimum size=0.15cm] (3) at (-1.5,1.5) {3};
    \node[draw, circle, inner sep=1pt,minimum size=0.15cm] (4) at (-1.5,-1.5) {4};
    \node[draw, circle, inner sep=1pt,minimum size=0.15cm] (5) at (1.5,-1.5) {5};
    \node[draw, circle, inner sep=1pt,minimum size=0.15cm] (6) at (0.75,0.75) {6};
    \node[draw, circle, inner sep=1pt,minimum size=0.15cm] (7) at (-0.75,0.75) {7};
    \path [] (3) edge node[] {} (7);
    \path [] (2) edge node[] {} (6);
    \path [] (7) edge node[] {} (6);
    \path [] (4) edge node[] {} (7);
    \path [] (5) edge node[] {} (6);
    \end{tikzpicture}
\end{minipage}
\end{center}
\caption{R3D's procedure (a)-(d) of duplicating nodes to enforce that all nodes have a valence of three.  (a) Vertex 1 is identified to be a vertex with valence of four.  (b) Vertex 1 is replaced with a cycle composing of four vertices 6, 7, 8, and 9 with coordinates equivalent to that of vertex 1.  (c) Two edge connected vertices 8 and 9 from the cycle are removed.  (d) The vertex clockwise from 8 in the cycle (7) is joined with 8's old neighbor (4).  The vertex counter-clockwise from 9 in the cycle (6) is joined with 9's old neighbor (5).}
\end{figure}
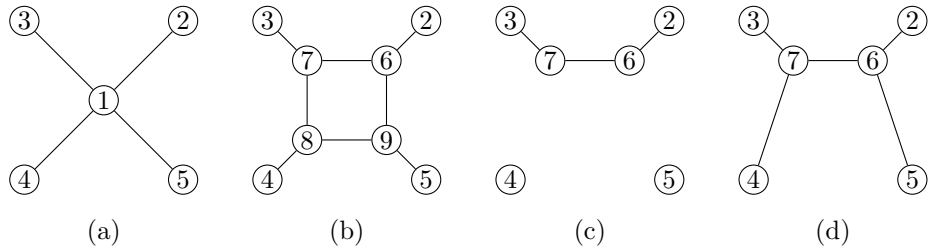

\subsection{R3D's Data Structures}
R3D avoids dynamic memory allocation by setting the maximum number of vertices via the compile time variable \texttt{R3D\_MAX\_VERTS} and by setting the number of neighbors per vertex to three.  Since R3D enforces that all vertices have three neighbors, this causes modification of the original polyhedron and sometimes a dramatic increase in the number of vertices.  As a consequence, \texttt{R3D\_MAX\_VERTS} has to be set to a very large size. 

\begin{figure}[H]
\begin{verbatim}
typedef struct
{
  r3d_int pnbrs[3]; /*!< Neighbor indices. */
  r3d_rvec3 pos;    /*!< Vertex position. */
} r3d_vertex;

typedef struct
{
  r3d_vertex verts[R3D_MAX_VERTS]; /*!< Vertex buffer. */
  r3d_int nverts;                  /*!< Number of vertices in the buffer. */
} r3d_poly;  
\end{verbatim}
\caption{The \texttt{r3d\_vertex} and \texttt{r3d\_poly} structs respectively used to represent a polyhedra in R3D.  Note that the \texttt{r3d\_vertex} has a buffer of size 3 to store neighbors.}
\end{figure}

\section{J3D}
We present a revised implementation of the Sugihara clip-n-cap algorithm that aims to remove some of the restrictions of R3D with an eye to improved performance on conventional CPUs and porting the algorithm to modern GPUs.  We introduce new data structures suited for this purpose and describe new algorithms to take advantage of these constructs.  

Given a face from a BREP and the ordered set of nodes forming that face, a directed edge is a pair of sequential nodes.  From this, the data structures we introduce are two maps $\mathcal{P}$ and $\mathcal{N}$.  These maps serve the dual purpose of cycling through the vertices of a face as well a cycling through the neighbors of a graph vertex, both in an ordered manner.  $\mathcal{P}$ maps a directed edge to the vertex preceding it in the face walk and $\mathcal{N}$ maps a directed edge to the vertex succeeding it in the face walk.  Each of these maps are represented as a $|V|\times |V|$ array of vertex indices where $V$ is the set of vertices in the polyhedron.
\begin{figure}[H]
\begin{center}
\begin{tikzpicture}
    \node[draw, circle, inner sep=1pt,minimum size=0.15cm] (1) at (0,0) {1};
    \node[draw, circle, inner sep=1pt,minimum size=0.15cm] (2) at (1.5,0) {2};
    \node[draw, circle, inner sep=1pt,minimum size=0.15cm] (3) at (1.5,1.5) {3};
    \node[draw, circle, inner sep=1pt,minimum size=0.15cm] (4) at (0,1.5) {4};

    \path [] (1) edge node[] {} (2);
    \path [] (2) edge node[] {} (3);
    \path [] (3) edge node[] {} (4);
    \path [] (4) edge node[] {} (1);

    \draw[ultra thick, ->] (0.75,0.1) arc (-90:170:0.6);
\end{tikzpicture}
\end{center}
$$\mathcal{P}(1,2)=4,\;\mathcal{P}(2,3)=1,\;\mathcal{P}(3,4)=2,\;\mathcal{P}(4,1)=3$$
$$\mathcal{N}(1,2)=3,\;\mathcal{N}(2,3)=4,\;\mathcal{N}(3,4)=1,\;\mathcal{N}(4,1)=2$$
\caption{An oriented four sided face and the corresponding maps $\mathcal{P}$ and $\mathcal{N}$.}
\end{figure}
Alternatively, $\mathcal{P}(A,B)$ can be viewed as mapping to the node following $B$ in the counter-clockwise iteration of neighbors around $A$.  In addition, $\mathcal{N}(A,B)$ can alse be viewed as mapping to the node preceding $A$ in the clockwise iteration of neighbors around $B$.

\begin{figure}[H]
\begin{center}
\begin{tikzpicture}
    \node[draw, circle, inner sep=1pt,minimum size=0.15cm] (1) at (0,0) {1};
    \node[draw, circle, inner sep=1pt,minimum size=0.15cm] (2) at (-1,-1) {2};
    \node[draw, circle, inner sep=1pt,minimum size=0.15cm] (3) at (1,-1) {3};
    \node[draw, circle, inner sep=1pt,minimum size=0.15cm] (4) at (1, 1) {4};
    \node[draw, circle, inner sep=1pt,minimum size=0.15cm] (5) at (-1, 1) {5};

    \path [] (1) edge node[] {} (2); 
    \path [] (1) edge node[] {} (3);
    \path [] (1) edge node[] {} (4);
    \path [] (1) edge node[] {} (5);
\end{tikzpicture}

\end{center}
$$\mathcal{P}(1,2)=3,\;\mathcal{P}(1,3)=4,\;\mathcal{P}(1,4)=5,\;\mathcal{P}(1,5)=2$$
$$\mathcal{N}(3,1)=2,\;\mathcal{N}(4,1)=3,\;\mathcal{N}(5,1)=4,\;\mathcal{N}(2,1)=5$$
\caption{Vertex 1, its neighbors, and the corresponding maps $\mathcal{P}$ and $\mathcal{N}$.}
\end{figure}

\subsection{Graph Initialization}

For initializing the graph from a BREP, we also define a map $\mathcal{S}:V\rightarrow V$ which for a vertex $v$ simply returns a neighbor of $v$.  This map, along with $\mathcal{P}$, is used to initialize the graph representation.  First, the algorithm to initialize $\mathcal{S}$ and $\mathcal{P}$ is as follows.  

\begin{algorithm}[H]
\caption{Initialization of the Maps $\mathcal{S}$ and $\mathcal{P}$}
\begin{algorithmic}
 \Require BREP representation $B$.
 \Ensure $\mathcal{S}$ and $\mathcal{P}$ are initialized.
 \Function{InitializationSetup}{$B$, $\mathcal{S}$, $\mathcal{P}$}
 \State $prev\leftarrow 0$
 \State $curr\leftarrow 0$
 \State $next\leftarrow 0$
 \For {$f\in B.faces$}
   \For {$i=1$ : $f.num\_vertices$}
     \State $prev\leftarrow f.vertices[(i-1)\mod f.num\_vertices]$
     \State $curr\leftarrow f.vertices[i]$
     \State $next\leftarrow f.vertices[(i+1)\mod f.num\_vertices]$
     \State $\mathcal{S}[curr]\leftarrow next$ 
     \State $\mathcal{P}[curr][next]\leftarrow prev$
   \EndFor
 \EndFor
 \EndFunction
\end{algorithmic}
\end{algorithm}

From here, the algorithm to initialize the graph with counter-clockwise vertex neighbor ordering is as follows.

\begin{algorithm}[H]
\caption{Counter-Clockwise Neighbors Initialization}
\begin{algorithmic}
 \Require $\mathcal{P}$, $\mathcal{S}$.
 \Ensure Graph $G$ with counter-clockwise neighbor ordering.
 \Function{NeighborsInitialization}{$G$, $\mathcal{P}$, $\mathcal{S}$}
 \For {$v\in G.vertices$}
   \State $G.neighbors[v].append(\mathcal{S}(v))$
   \State $v_{prev}\leftarrow \mathcal{P}(v, \mathcal{S}(v))$
   \While {$v_{prev}\neq \mathcal{S}(v)$} 
     \State $G.neighbors[v].append(v_{prev})$
     \State $v_{prev}\leftarrow \mathcal{P}(v, v_{prev})$
   \EndWhile
 \EndFor
 \EndFunction
\end{algorithmic}
\end{algorithm}

\subsection{Clip-N-Cap}

Before describing the clip-n-cap implementation in J3D, let's recall some details of the Sugihara algorithm from section 2.  When a plane clips a graph $G=(V,E)$, $V$ is partitioned into two sets $V_{1}$ and $V_{2}$.  $V_{1}$ is the set of vertices on the positive side of the plane and $V_{2}$ is the set of vertices on the negative side of the plane.  Our goal is to reconstruct the polyhedron on the positive side of the plane.  First, new vertices are generated on each edge connecting a vertex from $V_{1}$ to a vertex from $V_{2}$.  Now a detail which is important to mention is that after this step, clip-n-cap in R3D and J3D is implemented such that for each new vertex, the edge connecting it to a vertex in $V_{2}$ is deleted and the edge connecting it to a vertex in $V_{1}$ is kept.  Moreover, at this point, each new vertex has a singular neighbor which is a vertex in $V_{1}$.  This singular neighbor is the first element in the list of neighbors of a new vertex.  Before being able to execute clip-n-cap efficiently, we need to initialize $\mathcal{N}$ so we can walk from one new vertex to the next efficiently along a path of vertices in $V_{1}$.  So, we will initialize $\mathcal{N}$ by visiting each neighbor of each vertex in $V_{1}$.  We will take advantage of the fact that the neighbors of each vertex in $V_{1}$ are ordered counter clockwise relative to the exterior of the polyhedron.

\begin{algorithm}[H]
\caption{Initialization of the Map $\mathcal{N}$ After Initializing $V_{1}$, $V_{2}$, and New Vertices}
\begin{algorithmic}
 \Require Graph $G$ with a subset $V_{1}$ of the vertices with connectivity to the new vertices.
 \Ensure $\mathcal{N}$ is initialized.
 \Function{CapSetup}{$G$}
 \State $prev\leftarrow 0$
 \State $next\leftarrow 0$
 \For {$v\in V_{1}$}
   \For {$i=1$ : $G.num\_neighbors[v]$}
     \State $prev\leftarrow v.neighbors[i]$
     \State $next\leftarrow v.neighbors[(i - 1)\mod v.num\_neighbors]$
     \State $\mathcal{N}[prev][v]\leftarrow next$  
   \EndFor
 \EndFor
 \EndFunction
\end{algorithmic}
\end{algorithm}

Next, the algorithm to execute step two in Sugihara's clip-n-cap algorithm with the help of $\mathcal{N}$ is as follows.        

\begin{algorithm}[H]
\caption{Generating a Cycle Separating $V_{1}$ and $V_{2}$}
\begin{algorithmic}
  \Require $\mathcal{N}$
  \Require Graph $G$ has a set of new vertices $G.new\_vertices\subset V$ s.t. each new vertex has a singular neighbor in $V_{1}$.
  \Ensure The cycle of new vertices separating $V_{1}$ and $V_{2}$ is generated. 
  \Function{Cap}{$G$, $\mathcal{N}$}
  \For{$v_{new}\in G.new\_vertices$}
    \State $prev\leftarrow v_{new}$
    \State $curr\leftarrow prev.neighbors[0]$
    \State $next\leftarrow \mathcal{N}(prev, curr)$
    \While{$next\notin G.new\_vertices$}
      \State $prev\leftarrow curr$
      \State $curr\leftarrow next$
      \State $next\leftarrow \mathcal{N}(prev, curr)$
    \EndWhile
    \State $v_{new}.neighbors[1]\leftarrow next$
    \State $next.neighbors[2]\leftarrow v_{new}$
  \EndFor
  \EndFunction
\end{algorithmic}
\end{algorithm}

\subsection{BREP Traversal for Moment Calculation}
The map $\mathcal{N}$ will also prove necessary for efficient moments computation.  To compute the volume or the zeroth order moment, for example, we traverse each face and construct triangular facets from a starting vertex on the face.  Like in Figure 2, we begin with a starting vertex $A$ and the next two vertices $B$ and $C$ around the face of a polyhedron to form the first triangular facet.  So, in terms of the map $\mathcal{N}$ we defined earlier, we have that $\mathcal{N}(A,B)=C$ and the next facet we construct will be defined by the points $A$, $C$, and $\mathcal{N}(B,C)$ and so on.  This high level idea is leveraged by J3D to efficiently compute the moments of a polyhedron.  Before computing moments however, we need to initialize the map $\mathcal{N}$ as follows.
This algorithm is almost identical to how we initialized $\mathcal{N}$ before clip-n-cap except we consider all vertices in the polyhedron rather than just the vertices in $V_{1}$. 

\begin{algorithm}[H]
\caption{Initialization of the Map $\mathcal{N}$}
\begin{algorithmic}
 \Require Graph $G$ representing the polyhedron with neighbors around each vertex ordered ccw relative to the exterior 
 \Ensure $\mathcal{N}$ is initialized.
 \Function{VolumeComputationSetup}{$G$}
 \State $prev\leftarrow 0$
 \State $next\leftarrow 0$
 \For {$v\in G.vertices$}
   \For {$i=1$ : $G.num\_neighbors[v]$}
     \State $prev\leftarrow v.neighbors[i]$
     \State $next\leftarrow v.neighbors[(i - 1)\mod v.num\_neighbors]$
     \State $\mathcal{N}[prev][v]\leftarrow next$  
   \EndFor
 \EndFor
 \EndFunction
\end{algorithmic}
\end{algorithm}

By using the origin $O$ as a reference point, the signed volume of the tetrahedra formed by $O$, $A$, $B$, and $C$ is the following \cite{6127882}.
\begin{equation}
vol_{T}(A,B,C)=\frac{1}{6}
\begin{vmatrix}
x_{A} & y_{A} & z_{A}\\
x_{B} & y_{B} & z_{B}\\
x_{C} & y_{C} & z_{C}
\end{vmatrix}
\end{equation}
With this, the volume of the whole polyhedron can be computed via the following algorithm
\begin{algorithm}[H]
\caption{Volume Computation}
\begin{algorithmic}
 \Require $\mathcal{N}$
 \Require Graph $G$
 \Ensure Volume $vol$ of the polyhedra.
 \Function{BREPWalk}{$G$, $\mathcal{N}$, $vol$}
 \State $vol\leftarrow 0$
 \State $visited\_edges[G.num\_vertices][G.num\_vertices]$
 \For {$v\in G.vertices$}
   \For {$i=1$ : $G.num\_neighbors[v]$}
     \If{not $visited\_edges[v][G.neighbors[v][i]]$}
       \State $prev\leftarrow v$
       \State $curr\leftarrow G.neighbors[v][i]$
       \State $next\leftarrow \mathcal{N}(prev, curr)$
       \State $visited\_edges[prev][curr]\leftarrow true$
       \While{$next\neq v$}
         \State $vol\leftarrow vol + vol_{T}(v,curr,next)$
         \State $prev\leftarrow curr$
         \State $curr\leftarrow next$
         \State $next\leftarrow \mathcal{N}(prev, curr)$
         \State $visited\_edges[prev][curr]\leftarrow true$
       \EndWhile
       \State $visited\_edges[curr][next]\leftarrow true$
     \EndIf
   \EndFor
 \EndFor
 \EndFunction
\end{algorithmic}
\end{algorithm}

\subsection{J3D Implementation}
J3D is a C library which only uses static allocations for its working buffers.  However, since we have no restrictions on the valence of nodes, J3D has two compile time variables \texttt{J3D\_MAX\_VERTS} and \texttt{J3D\_MAX\_EDGES} to statically size its working buffers.  Since the graph representing a polyhedra is always isomorphic to a planar graph and is therefore sparse, \texttt{J3D\_MAX\_EDGES} can be set to a positive integer multiple of \texttt{J3D\_MAX\_VERTS}.  The adjacency of the graph is stored in a compressed sparse rows (CSR) array \cite{CSR}.  That is, rather than storing the neighbors of an vertex in an individual array for each vertex, neighbors are stored in a contiguous chunk of a larger array and a separate row index array is used to index into the correct chunk of a vertex. 

\begin{figure}[H]
\begin{verbatim}
typedef struct {
  double verts[J3D_MAX_VERTS][3];   /*!< Vertex positions. */
  int dir_edges[2 * J3D_MAX_EDGES]; /*!< CSR array of the graph adjacency. */
  int indices[J3D_MAX_VERTS + 1];   /*!< Offsets for the CSR array. */
  int num_dir_edges;                /*!< Number of directed edges. */
  int num_verts;                    /*!< Number of vertices. */
} j3d_poly;
\end{verbatim}
\caption{The struct used to represent a polyhedron in J3D.  \texttt{J3D\_MAX\_EDGES} is used to reserve a fixed sixed buffer for a CSR array used to store the graph adjacency.  \texttt{J3D\_MAX\_VERTS} is used to reserve fixed sized buffers for the vertex positions and the row indices for the CSR array.}
\end{figure}

\begin{center}
\begin{tikzpicture}
    \node[draw, circle, inner sep=1pt,minimum size=0.15cm] (1) at (1.65,2.5) {1};
    \node[draw, circle, inner sep=1pt,minimum size=0.15cm] (2) at (0,0) {2};
    \node[draw, circle, inner sep=1pt,minimum size=0.15cm] (3) at (2,0) {3};
    \node[draw, circle, inner sep=1pt,minimum size=0.15cm] (4) at (3.3,1) {4};
    \node[draw, circle, inner sep=1pt,minimum size=0.15cm] (5) at (1.3,1) {5};
    
    \path [] (2) edge node[] {} (3); 
    \path [] (3) edge node[] {} (4);
    \path [dashed] (4) edge node[] {} (5); 
    \path [dashed] (5) edge node[] {} (2); 
    
    \path [] (2) edge node[] {} (1);
    \path [] (3) edge node[] {} (1);
    \path [] (4) edge node[] {} (1);
    \path [dashed] (5) edge node[] {} (1);
\end{tikzpicture}
\end{center}

\begin{figure}[h]
\centering
\begin{tikzpicture}[square/.style={draw,
                                   align=center,
                                   text width=0.4cm,
                                   minimum width=0.4cm,
                                   minimum height=0.4cm}]
    \node[square] (0) at (0,0) {2};
    \node[square] (1) at (0.64,0) {3};
    \node[square] (2) at (1.28,0) {4};
    \node[square] (3) at (1.92,0) {5};
    \node[square] (4) at (2.56,0) {3};
    \node[square] (5) at (3.2,0) {1};
    \node[square] (6) at (3.84,0) {5};
    \node[square] (7) at (4.48,0) {4};
    \node[square] (8) at (5.12,0) {1};
    \node[square] (9) at (5.76,0) {2};
    \node[square] (10) at (6.4,0) {5};
    \node[square] (11) at (7.04,0) {1};
    \node[square] (12) at (7.68,0) {3};
    \node[square] (13) at (8.32,0) {2};
    \node[square] (14) at (8.96,0) {1};
    \node[square] (15) at (9.6,0) {4};
    \node[square] (16) at (10.24,0) {\textcolor{white}{0}};

    \node[square] (17) at (3.2,-1.5) {0};
    \node[square] (18) at (3.84,-1.5) {4};
    \node[square] (19) at (4.48,-1.5) {7};
    \node[square] (20) at (5.12,-1.5) {10};
    \node[square] (21) at (5.76,-1.5) {13};
    \node[square] (22) at (6.4,-1.5) {16};

    \path [->] (17.north) edge node[] {} (0.south);
    \path [->] (18.north) edge node[] {} (4.south);
    \path [->] (19.north) edge node[] {} (7.south);
    \path [->] (20.north) edge node[] {} (10.south);
    \path [->] (21.north) edge node[] {} (13.south);
    \path [->] (22.north) edge node[] {} (16.south);

    \draw [decorate,decoration={brace,amplitude=5pt,raise=2ex},align=center]
  (-0.3,0.2) -- (2.22,0.2) node[midway,yshift=2.5em]{\footnotesize{neighbors of}\\\footnotesize{vertex 1}};
    \draw [decorate,decoration={brace,amplitude=5pt,raise=2ex},align=center]
  (2.26,0.2) -- (4.14,0.2) node[midway,yshift=2.5em]{\footnotesize{neighbors of}\\\footnotesize{vertex 2}};
    \draw [decorate,decoration={brace,amplitude=5pt,raise=2ex},align=center]
  (4.18,0.2) -- (6.06,0.2) node[midway,yshift=2.5em]{\footnotesize{neighbors of}\\\footnotesize{vertex 3}};
    \draw [decorate,decoration={brace,amplitude=5pt,raise=2ex},align=center]
  (6.10,0.2) -- (7.98,0.2) node[midway,yshift=2.5em]{\footnotesize{neighbors of}\\\footnotesize{vertex 4}};
    \draw [decorate,decoration={brace,amplitude=5pt,raise=2ex},align=center]
  (8.02,0.2) -- (9.9,0.2) node[midway,yshift=2.5em]{\footnotesize{neighbors of}\\\footnotesize{vertex 5}};
\end{tikzpicture}
\caption{A quadrilateral pyramid with its CSR array representation.  The first array is used to store the adjacency or all the neighbors contiguously.  Moreover, the second array consists of indices used to index into the neighbors of each vertex in the first array.  There is an additional element at the end of the second array so that we know when the contents of the first array end.}
\end{figure}
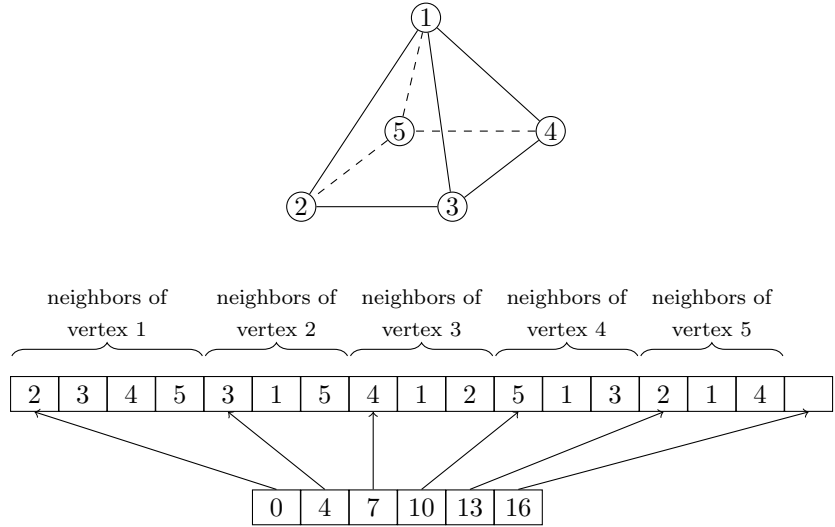

\subsection{Additional Implementation Improvements}  
In moments calculation, J3D explicitly computes the zeroth, first, second, and third order moments and falls back to the recursive method in \cite{6127882} for higher order moments whereas R3D uses the recursive method for all moments.  The explicit formulas for the moments computation can be found in Appendix A.

\subsection{Within The Context of Modern Day Computing}
R3D was implemented primarily with CPU computation in mind but did make some optimizations for vectorization. Since its development, the computational power of new supercomputers is coming primarily from GPUs (over 95\%) while the memory capacity of both CPUs and GPUs is still constrained.  J3D is optimized on both counts for the GPU. Given that J3D polyhedra themselves are facing size constraints on modern GPUs, the explosion in size of R3D polygons coming from the trivalent requirement is a non-starter for such platforms.

\section{Lessons Learned}
It is possible to implement a data structure representing a graph with arbitrary valence utilizing a \texttt{J3D\_MAX\_DEGREE} compile variable time instead.  Such variable is used to statically allocate a working buffer for each vertex to store its neighbors.  While this improves the performance on small scale benchmarks, it ended up being slower when integrated into a larger code.
\begin{figure}[H]
\begin{verbatim}
typedef struct
{
  int pnbrs[J3D_MAX_DEGREE]; /*!< Neighbor indices. */
  int degree;                /*!< Number of neighbors. */
  j3d_vec3 pos;              /*!< Vertex position. */
} j3d_vertex;

typedef struct
{
  j3d_vertex verts[J3D_MAX_VERTS]; /*!< Vertex buffer. */
  int nverts;                      /*!< Number of vertices in the buffer. */
} j3d_poly;  
\end{verbatim}
\caption{The \texttt{j3d\_vertex} and \texttt{j3d\_poly} structs respectively used to represent a polyhedron with a \texttt{J3D\_MAX\_DEGREE} compile time variable.}
\end{figure}

\begin{center}
\begin{tikzpicture}
    \node[draw,circle,inner sep=1pt,minimum size=0.15cm] (1) at (1.65,2.5) {1};
    \node[draw,circle,inner sep=1pt,minimum size=0.15cm] (2) at (0.2,0.1) {2};
    \node[draw,circle,inner sep=1pt,minimum size=0.15cm] (3) at (1.7,0.1) {3};
    \node[draw,circle,inner sep=1pt,minimum size=0.15cm] (4) at (2.8,0.5) {4};
    \node[draw,circle,inner sep=1pt,minimum size=0.15cm] (5) at (2.1,1) {5};
    \node[draw,circle,inner sep=1pt,minimum size=0.15cm] (6) at (1.0,1) {6};
    \node[draw,circle,inner sep=1pt,minimum size=0.15cm] (7) at (-0.1,0.7) {7};

    \path [] (2) edge node[] {} (3);
    \path [] (3) edge node[] {} (4);
    \path [dashed] (4) edge node[] {} (5);
    \path [dashed] (5) edge node[] {} (6);
    \path [dashed] (6) edge node[] {} (7);
    \path [] (7) edge node[] {} (2);
    
    \path [] (2) edge node[] {} (1);
    \path [] (3) edge node[] {} (1);
    \path [] (4) edge node[] {} (1);
    \path [dashed] (5) edge node[] {} (1);
    \path [dashed] (6) edge node[] {} (1);
    \path [] (7) edge node[] {} (1);
\end{tikzpicture}
\end{center}

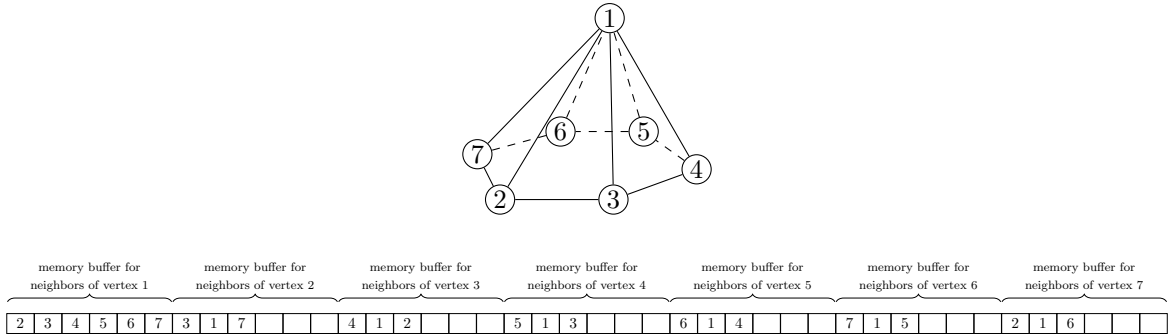
\begin{figure}[h]
\centering
\resizebox{\textwidth}{!}{
\begin{tikzpicture}[square/.style={draw,
                                   align=center,
                                   text width=0.4cm,
                                   minimum width=0.4cm,
                                   minimum height=0.4cm}]

    \node[square] (0) at (0.0,0) {2};
    \node[square] (1) at (0.64,0) {3};
    \node[square] (2) at (1.28,0) {4};
    \node[square] (3) at (1.92,0) {5};
    \node[square] (4) at (2.56,0) {6};
    \node[square] (5) at (3.2,0) {7};
    
    \node[square] (6) at (3.84,0) {3};
    \node[square] (7) at (4.48,0) {1};
    \node[square] (8) at (5.12,0) {7};
    \node[square] (9) at (5.76,0) {\textcolor{white}{0}};
    \node[square] (10) at (6.4,0) {\textcolor{white}{0}};
    \node[square] (11) at (7.04,0) {\textcolor{white}{0}};
    
    \node[square] (12) at (7.68,0) {4};
    \node[square] (13) at (8.32,0) {1};
    \node[square] (14) at (8.96,0) {2};
    \node[square] (15) at (9.6,0) {\textcolor{white}{0}};
    \node[square] (16) at (10.24,0) {\textcolor{white}{0}};
    \node[square] (17) at (10.88,0) {\textcolor{white}{0}};
    
    \node[square] (18) at (11.52,0) {5};
    \node[square] (19) at (12.16,0) {1};
    \node[square] (20) at (12.8,0) {3};
    \node[square] (21) at (13.44,0) {\textcolor{white}{0}};
    \node[square] (22) at (14.08,0) {\textcolor{white}{0}};
    \node[square] (23) at (14.72,0) {\textcolor{white}{0}};
    
    \node[square] (24) at (15.36,0) {6};
    \node[square] (25) at (16.0,0) {1};
    \node[square] (26) at (16.64,0) {4};
    \node[square] (27) at (17.28,0) {\textcolor{white}{0}};
    \node[square] (28) at (17.92,0) {\textcolor{white}{0}};
    \node[square] (29) at (18.56,0) {\textcolor{white}{0}};
    
    \node[square] (30) at (19.2,0) {7};
    \node[square] (31) at (19.84,0) {1};
    \node[square] (32) at (20.48,0) {5};
    \node[square] (33) at (21.12,0) {\textcolor{white}{0}};
    \node[square] (34) at (21.76,0) {\textcolor{white}{0}};
    \node[square] (35) at (22.40,0) {\textcolor{white}{0}};
    
    \node[square] (36) at (23.04,0) {2};
    \node[square] (37) at (23.68,0) {1};
    \node[square] (38) at (24.32,0) {6};
    \node[square] (39) at (24.96,0) {\textcolor{white}{0}};
    \node[square] (40) at (25.6,0) {\textcolor{white}{0}};
    \node[square] (41) at (26.24,0) {\textcolor{white}{0}};

    \draw [decorate,decoration={brace,amplitude=5pt,raise=2ex},align=center]
    (-0.3,0.2) -- (3.5,0.2) node[midway,yshift=2.5em]{\footnotesize{memory buffer for }\\\footnotesize{neighbors of vertex 1}};

    \draw [decorate,decoration={brace,amplitude=5pt,raise=2ex},align=center]
    (3.53,0.2) -- (7.34,0.2) node[midway,yshift=2.5em]{\footnotesize{memory buffer for }\\\footnotesize{neighbors of vertex 2}};

    \draw [decorate,decoration={brace,amplitude=5pt,raise=2ex},align=center]
    (7.37,0.2) -- (11.18,0.2) node[midway,yshift=2.5em]{\footnotesize{memory buffer for }\\\footnotesize{neighbors of vertex 3}};

    \draw [decorate,decoration={brace,amplitude=5pt,raise=2ex},align=center]
    (11.21,0.2) -- (15.02,0.2) node[midway,yshift=2.5em]{\footnotesize{memory buffer for }\\\footnotesize{neighbors of vertex 4}};

    \draw [decorate,decoration={brace,amplitude=5pt,raise=2ex},align=center]
    (15.05,0.2) -- (18.86,0.2) node[midway,yshift=2.5em]{\footnotesize{memory buffer for }\\\footnotesize{neighbors of vertex 5}};

    \draw [decorate,decoration={brace,amplitude=5pt,raise=2ex},align=center]
    (18.89,0.2) -- (22.70,0.2) node[midway,yshift=2.5em]{\footnotesize{memory buffer for }\\\footnotesize{neighbors of vertex 6}};

    \draw [decorate,decoration={brace,amplitude=5pt,raise=2ex},align=center]
    (22.73,0.2) -- (26.54,0.2) node[midway,yshift=2.5em]{\footnotesize{memory buffer for }\\\footnotesize{neighbors of vertex 7}};
  
\end{tikzpicture}
}
\caption{A hexagonal pyramid with an overview of how its adjacency is represented in memory with \texttt{J3D\_MAX\_DEGREE=6}.  Empty memory cells represent unused or redundant memory cells.  As a consequence, the representation is memory inefficient and cache misses occur far more frequently than if the polyhedron is represented with a CSR array.}
\end{figure}

This performance loss was attributable to a lack of memory locality which led to far more cache misses.  Since the majority of vertices do not have \texttt{J3D\_MAX\_DEGREE} many neighbors, those vertices have unused information at the end of their respective buffer.  Consequently, the memory used to store the entire polyhedron is fragmented and cache misses inevitably arise.  This led to the implementation in section 4.4 which has optimal memory locality and superior performance across various tests as demonstrated in the next section.

\section{Benchmarks}
Our algorithm makes improvements to the serial implementation of a single clip operation on a polyhedron over R3D's implementation and therefore, we are doing serial comparisons.  While this clipping algorithm will indeed be used in parallel in real applications, we did not want to muddy the performance comparisons for the core algorithm.

\subsection{Kernel Level Benchmarks}
For the following kernel level benchmark, the hardware we utilized was the Intel(R) Xeon(R) CPU E5-2695 v4 @ 2.10GHz.  R3D and J3D were compiled with gcc 8.5.0.  The default buffer sizes for J3D and R3D were used, that is, \texttt{J3D\_MAX\_VERTS=256}, \texttt{J3D\_MAX\_EDGES=512}, and \texttt{R3D\_MAX\_VERTS=512}.  We retrieved the results from HPCToolkit \cite{HPCTK}.

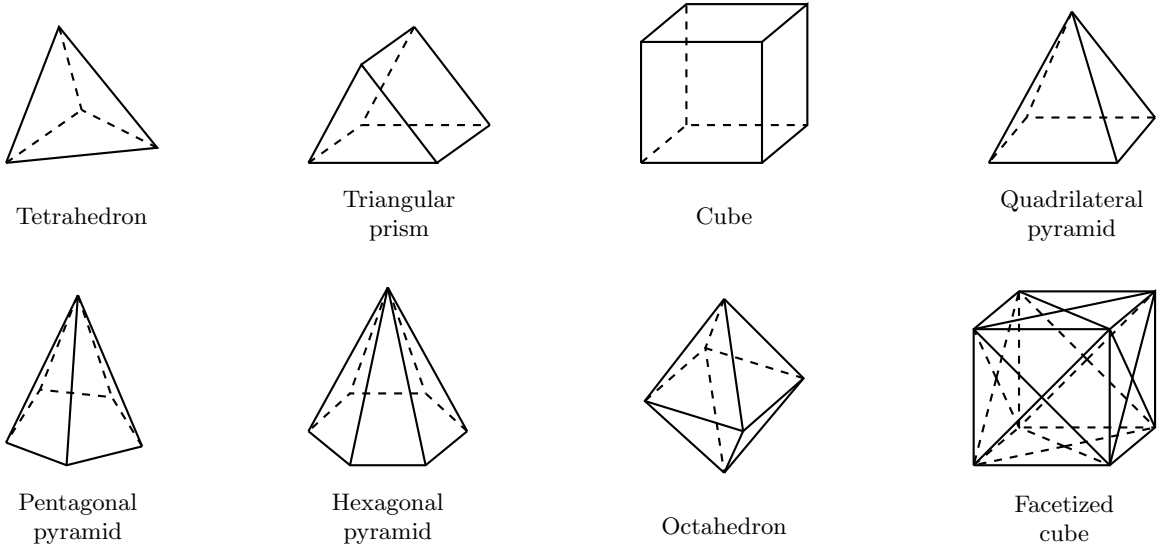
\begin{figure}[H]
    \centering
    \begin{tikzpicture}[
    scale=1.0,
    vertex/.style={circle, fill=black, inner sep=1.1pt},
    edge/.style={thick},
    hidden/.style={thick, dashed},
    label/.style={font=\small, align=center}
]


\begin{scope}[shift={(0,0)}]
    \coordinate (A) at (0,0);
    \coordinate (B) at (2,0.2);
    \coordinate (C) at (0.7,1.8);
    \coordinate (D) at (1.0,0.7);

    \draw[edge] (A)--(B)--(C)--(A);

    \draw[hidden] (A)--(D);
    \draw[hidden] (B)--(D);
    \draw[hidden] (C)--(D);

    \node[label] at (1,-0.7) {Tetrahedron};
\end{scope}

\% 2. Triangular prism
\begin{scope}[shift={(4,0)}]
    \coordinate (A) at (0,0);
    \coordinate (B) at (1.7,0);
    \coordinate (C) at (0.7,1.3);
    \coordinate (Ap) at (0.7,0.5);
    \coordinate (Bp) at (2.4,0.5);
    \coordinate (Cp) at (1.4,1.8);

    \draw[edge] (A)--(B)--(C)--(A);

    \draw[hidden] (Ap)--(Bp);
    \draw[hidden] (Cp)--(Ap);
    \draw[edge] (Bp)--(Cp);

    \draw[hidden] (A)--(Ap);
    \draw[edge] (B)--(Bp);
    \draw[edge] (C)--(Cp);

    \node[label] at (1.2,-0.7) {Triangular\\prism};
\end{scope}

\begin{scope}[shift={(8.4,0)}]
    \coordinate (A) at (0,0);
    \coordinate (B) at (1.6,0);
    \coordinate (C) at (1.6,1.6);
    \coordinate (D) at (0,1.6);
    \coordinate (E) at (0.6,0.5);
    \coordinate (F) at (2.2,0.5);
    \coordinate (G) at (2.2,2.1);
    \coordinate (H) at (0.6,2.1);

    \draw[edge] (A)--(B)--(C)--(D)--(A);
    \draw[edge] (B)--(F)--(G)--(C);
    \draw[edge] (D)--(H)--(G);
    \draw[hidden] (A)--(E)--(F);
    \draw[hidden] (E)--(H);

    \node[label] at (1.1,-0.7) {Cube};
\end{scope}

\begin{scope}[shift={(13,0)}]
    \coordinate (A) at (0,0);
    \coordinate (B) at (1.7,0);
    \coordinate (C) at (2.2,0.6);
    \coordinate (D) at (0.5,0.6);
    \coordinate (S) at (1.1,2.0);

    \draw[edge] (A)--(B)--(C);
    \draw[hidden] (C)--(D)--(A);
    \draw[edge] (S)--(A);
    \draw[edge] (S)--(B);
    \draw[edge] (S)--(C);
    \draw[hidden] (S)--(D);

    \node[label] at (1.1,-0.7) {Quadrilateral\\pyramid};
\end{scope}


\begin{scope}[shift={(0,-4)}]
    \coordinate (A) at (0,0.3);
    \coordinate (B) at (0.8,0);
    \coordinate (C) at (1.8,0.25);
    \coordinate (D) at (1.4,0.9);
    \coordinate (E) at (0.45,1.0);
    \coordinate (S) at (0.95,2.25);

    \draw[edge] (A)--(B)--(C);
    \draw[hidden] (C)--(D)--(E)--(A);
    \draw[edge] (S)--(A);
    \draw[edge] (S)--(B);
    \draw[edge] (S)--(C);
    \draw[hidden] (S)--(D);
    \draw[hidden] (S)--(E);

    \node[label] at (0.95,-0.7) {Pentagonal\\pyramid};
\end{scope}

\begin{scope}[shift={(4,-4)}]
    \coordinate (A) at (0,0.45);
    \coordinate (B) at (0.55,0);
    \coordinate (C) at (1.55,0);
    \coordinate (D) at (2.1,0.45);
    \coordinate (E) at (1.55,0.95);
    \coordinate (F) at (0.55,0.95);
    \coordinate (S) at (1.05,2.35);

    \draw[edge] (A)--(B)--(C)--(D);
    \draw[hidden] (D)--(E)--(F)--(A);
    \draw[edge] (S)--(A);
    \draw[edge] (S)--(B);
    \draw[edge] (S)--(C);
    \draw[edge] (S)--(D);
    \draw[hidden] (S)--(E);
    \draw[hidden] (S)--(F);

    \node[label] at (1.05,-0.7) {Hexagonal\\pyramid};
\end{scope}

\begin{scope}[shift={(8.5,-4)}]
    \coordinate (T) at (1,2.2);
    \coordinate (B) at (1,-0.1);

    \coordinate (L) at (-0.05,0.85);
    \coordinate (R) at (2.05,1.15);
    \coordinate (F) at (1.25,0.45);
    \coordinate (K) at (0.75,1.55);

    \draw[edge] (T)--(L)--(B);
    \draw[edge] (T)--(F)--(B);
    \draw[edge] (T)--(R)--(B);
    \draw[edge] (L)--(F)--(R);

    \draw[hidden] (T)--(K)--(B);
    \draw[hidden] (L)--(K)--(R);

    \node[label] at (1,-0.8) {Octahedron};
\end{scope}

\begin{scope}[shift={(12.8,-4)}]
    \coordinate (A) at (0,0);       
    \coordinate (B) at (1.8,0);     
    \coordinate (C) at (1.8,1.8);   
    \coordinate (D) at (0,1.8);     

    \coordinate (E) at (0.6,0.5);   
    \coordinate (F) at (2.4,0.5);   
    \coordinate (G) at (2.4,2.3);   
    \coordinate (H) at (0.6,2.3);   

    \coordinate (Ofront)  at (0.9,0.9);    
    \coordinate (Oback)   at (1.5,1.4);    
    \coordinate (Oright)  at (2.1,1.15);   
    \coordinate (Oleft)   at (0.3,1.15);   
    \coordinate (Otop)    at (1.2,2.05);   
    \coordinate (Obottom) at (1.2,0.25);   

    \draw[edge] (A)--(B)--(C)--(D)--(A);
    \draw[edge] (B)--(F)--(G)--(C);
    \draw[edge] (D)--(H)--(G);

    \draw[hidden] (A)--(E)--(F);
    \draw[hidden] (E)--(H);

    \draw[edge] (Ofront)--(A);
    \draw[edge] (Ofront)--(B);
    \draw[edge] (Ofront)--(C);
    \draw[edge] (Ofront)--(D);

    \draw[edge] (Oright)--(B);
    \draw[edge] (Oright)--(F);
    \draw[edge] (Oright)--(G);
    \draw[edge] (Oright)--(C);

    \draw[edge] (Otop)--(D);
    \draw[edge] (Otop)--(C);
    \draw[edge] (Otop)--(G);
    \draw[edge] (Otop)--(H);

    \draw[hidden] (Oback)--(E);
    \draw[hidden] (Oback)--(F);
    \draw[hidden] (Oback)--(G);
    \draw[hidden] (Oback)--(H);

    \draw[hidden] (Oleft)--(A);
    \draw[hidden] (Oleft)--(E);
    \draw[hidden] (Oleft)--(H);
    \draw[hidden] (Oleft)--(D);

    \draw[hidden] (Obottom)--(A);
    \draw[hidden] (Obottom)--(B);
    \draw[hidden] (Obottom)--(F);
    \draw[hidden] (Obottom)--(E);

    \node[label] at (1.2,-0.7) {Facetized\\cube};
\end{scope}
\end{tikzpicture}
\caption{The eight polyhedra we tested for the kernel level benchmarks.}
\end{figure}

We considered eight polyhedra starting in the BREP representation.  For each of these polyhedra, we ran $10^{7}$ iterations of initializing the polyhedron into the internal representation of R3D and J3D, conducting clip-n-cap with a random plane through the center, then computing the resulting volume.  A polyhedron we want to emphasize is the facetized cube, which is a cube with each of its faces decomposed into triangular faces (Figure 1).  This decomposition is a common practice used to decompose potentially curved faces of a polyhedral cell into planar faces \cite{AHN20072096}.  Another reason for using this polyhedron in our benchmarks is to show how expensive the initialization step of R3D is especially when the polyhedron has vertices with degree greater than three and the algorithm in Figure 4 must be invoked.

\pgfplotstableread[row sep=\\,col sep=&]{
 polyhedra             & J3D & R3D \\
 tetrahedron           & 2.80 & 4.43 \\
 triangular prism      & 3.57 & 5.12 \\
 cube                  & 4.38 & 5.79 \\
 quadrilateral pyramid & 3.44 & 11.7 \\
 pentagonal pyramid    & 3.44 & 11.3 \\
 hexagonal pyramid     & 3.92 & 15.3 \\
 octahedron            & 4.47 & 14.0 \\
 facetized cube     & 11.2 & 59.0 \\
    }\mydata

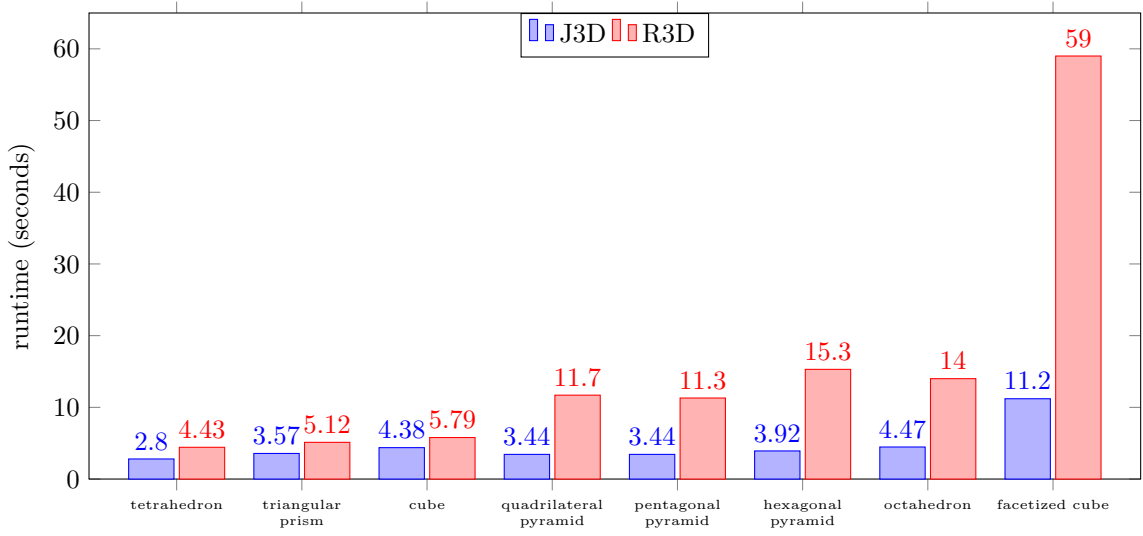
\begin{figure}[H]
\begin{tikzpicture}
    \begin{axis}[
            ybar,
            bar width=.6cm,
            width=\textwidth,
            height=.5\textwidth,
            legend style={at={(0.5,1)},
                anchor=north,legend columns=-1},
            symbolic x coords={tetrahedron,
                               triangular prism,
                               cube,
                               quadrilateral pyramid,
                               pentagonal pyramid,
                               hexagonal pyramid,
                               octahedron,
                               facetized cube},
            x tick label style={font=\tiny,text width=1.7cm,align=center},
            xtick=data,
            nodes near coords,
            nodes near coords align={vertical},
            ymin=0,ymax=65,
            ylabel={runtime (seconds)},
        ]
        \addplot table[x=polyhedra,y=J3D]{\mydata};
        \addplot table[x=polyhedra,y=R3D]{\mydata};
        \legend{J3D, R3D}
    \end{axis}
\end{tikzpicture}
\caption{Runtime in seconds for J3D and R3D to run $10^{7}$ iterations of initializing a polyhedron from the BREP to the internal data structure, conducting clip-n-cap with a random plane through the centroid, then computing the resulting volume.} 
\end{figure}

\pgfplotstableread[row sep=\\,col sep=&]{
 method                     & J3D & R3D \\
 initialization             & 3.33 & 46.2 \\
 clip-n-cap                       & 2.82 & 4.52 \\
 volume computation         & 4.76 & 7.99 \\
    }\mydata

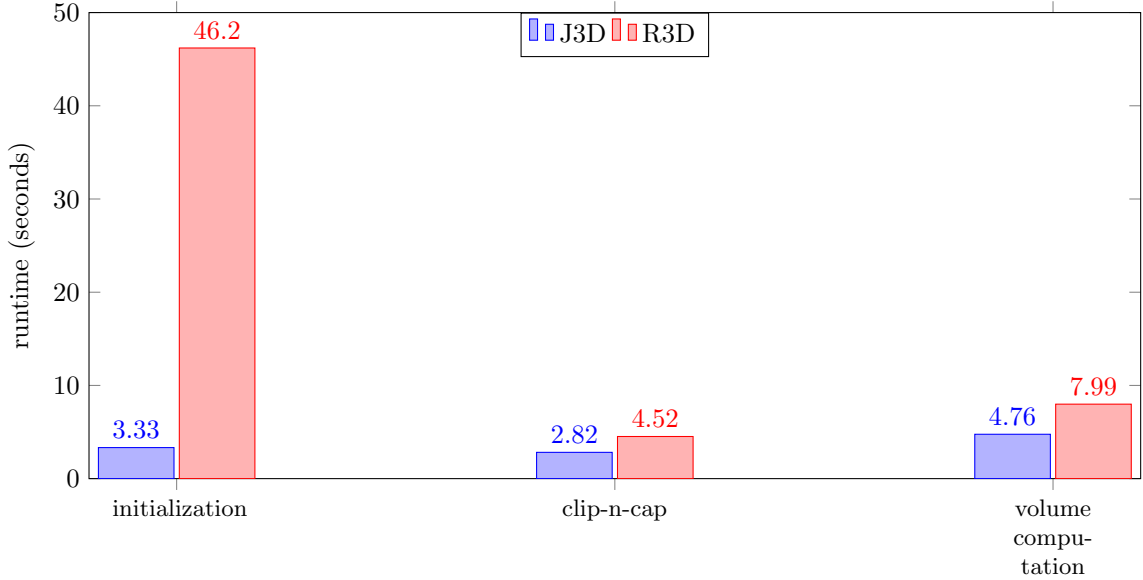
\begin{figure}[H]
\begin{tikzpicture}
    \begin{axis}[
            ybar,
            bar width=1cm,
            width=\textwidth,
            height=.5\textwidth,
            legend style={at={(0.5,1)},
                anchor=north,legend columns=-1},
            symbolic x coords={initialization,
                               clip-n-cap,
                               volume computation},
            x tick label style={font=\small,text width=1.7cm,align=center},
            xtick=data,
            nodes near coords,
            nodes near coords align={vertical},
            ymin=0,ymax=50,
            ylabel={runtime (seconds)},
        ]
        \addplot table[x=method,y=J3D]{\mydata};
        \addplot table[x=method,y=R3D]{\mydata};
        \legend{J3D, R3D}
    \end{axis}
\end{tikzpicture}
\caption{Runtime in seconds for each method in J3D and R3D after running $10^{7}$ iterations of initializing a facetized cube from the BREP to the internal data structure, conducting clip-n-cap with a random plane through the centroid, then computing the resulting volume.} 
\end{figure}

\subsection{Rotor Problem}
A problem in interface reconstruction that we benchmarked is the rotor problem.  In this problem, our material interface reconstruction code Tangram \cite{Herring-2021} is used to recover the material geometry of a complex series of intermeshed gears or rotors onto a 288x288x288 cartesian structured mesh.

\begin{figure}[h]
    \centering
    \includegraphics[width=0.5\textwidth]{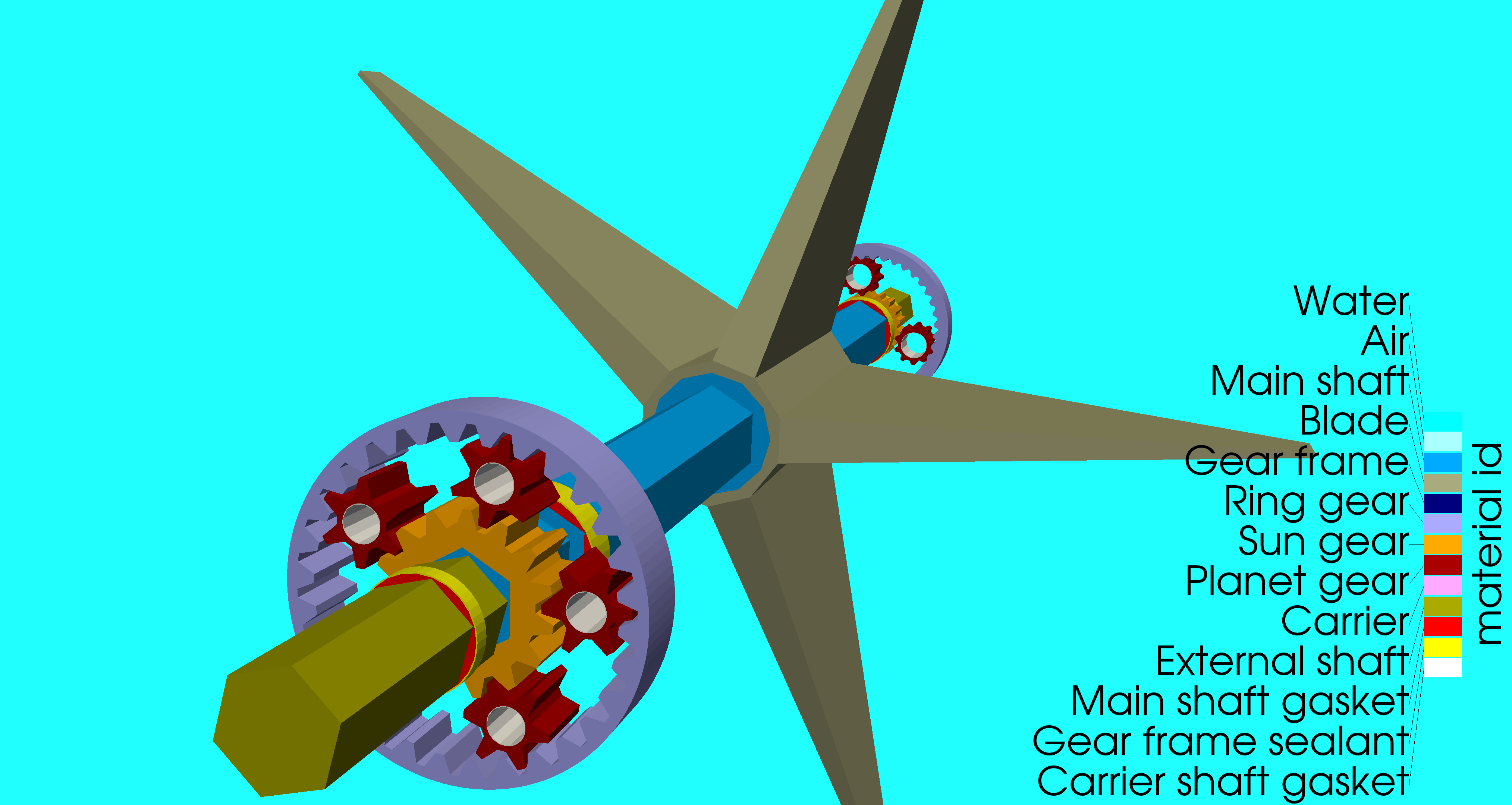}
    \caption{A picture of the reference material interfaces in the rotor problem demonstrating the scale and complexity of the problem.}
\end{figure}

In the following benchmarks, the hardware we utilized was the AMD EPYC 7513 32-Core Processor @ 2.60GHz.  Tangram and its dependencies were compiled with gcc/11.2.0.  The buffer sizes for J3D and R3D required by Tangram were used, that is, \texttt{J3D\_MAX\_VERTS=256}, \texttt{J3D\_MAX\_EDGES=512}, and \texttt{R3D\_MAX\_VERTS=1024}.  We ran several instances of the rotor problem with different material interface reconstruction strategies (VOF, MOF, LVIRA, and LVIRA+ which reconstruct the interfaces with varying degrees of fidelity) \cite{VOF, dyadechko2005moment, LVIRA, lviraplus}. 

\pgfplotstableread[row sep=\\,col sep=&]{
 method          & J3D & R3D \\
 VOF             & 5.54 & 5.13 \\
 MOF             & 64.4 & 168 \\
 LVIRA           & 44.7 & 94.5 \\
 LVIRA+       & 224 & 1109 \\
    }\mydata

\begin{figure}[H]
\begin{tikzpicture}
    \begin{axis}[
            ybar,
            bar width=1cm,
            width=\textwidth,
            height=.5\textwidth,
            legend style={at={(0.5,1)},
                anchor=north,legend columns=-1},
            symbolic x coords={VOF,
                               MOF,
                               LVIRA,
                               LVIRA+},
            x tick label style={font=\small,text width=1.7cm,align=center},
            xtick=data,
            nodes near coords,
            nodes near coords align={vertical},
            ymin=0,ymax=1200,
            ylabel={runtime (seconds)},
        ]
        \addplot table[x=method,y=J3D]{\mydata};
        \addplot table[x=method,y=R3D]{\mydata};
        \legend{J3D, R3D}
    \end{axis}
\end{tikzpicture}
\caption{Runtime in seconds for J3D and R3D with various interface reconstruction methods to run the rotor problem on a 288x288x288 cartesian structured mesh.} 
\end{figure}
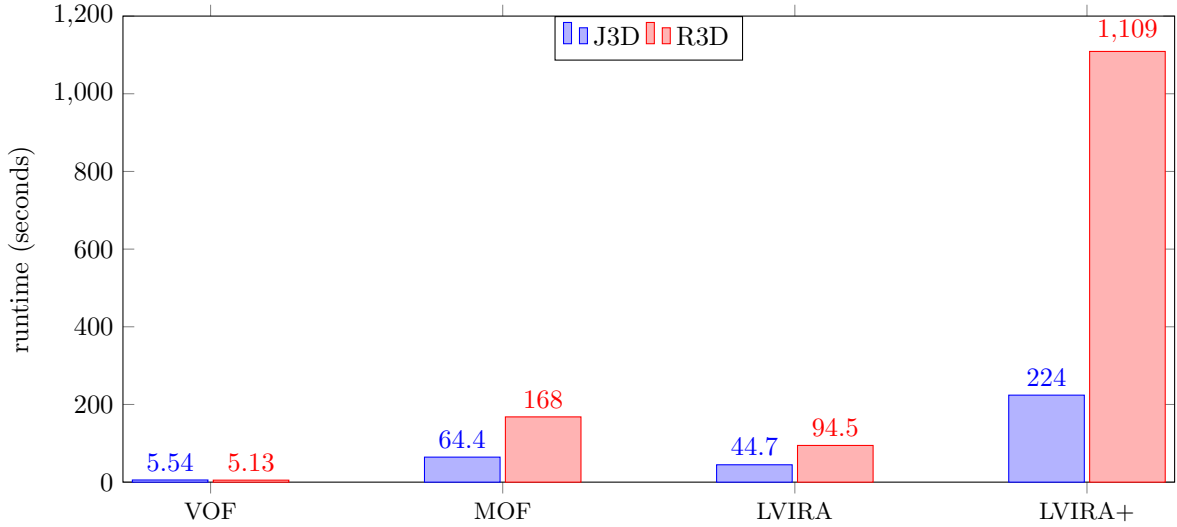

\subsection{Rocket Problem}
To demonstrate the performance improvements in mesh interpolation and material interface reconstruction, we also benchmarked the rocket problem.  In this problem, we begin with a boundary conformal mesh in the shape of a rocket.  We remap this mesh onto a structured mesh and remap this result onto a slightly shifted structured mesh to exercise remap and interface reconstruction in the same run.

\begin{figure}[h]
    \centering
    \includegraphics[width=0.4\textwidth]{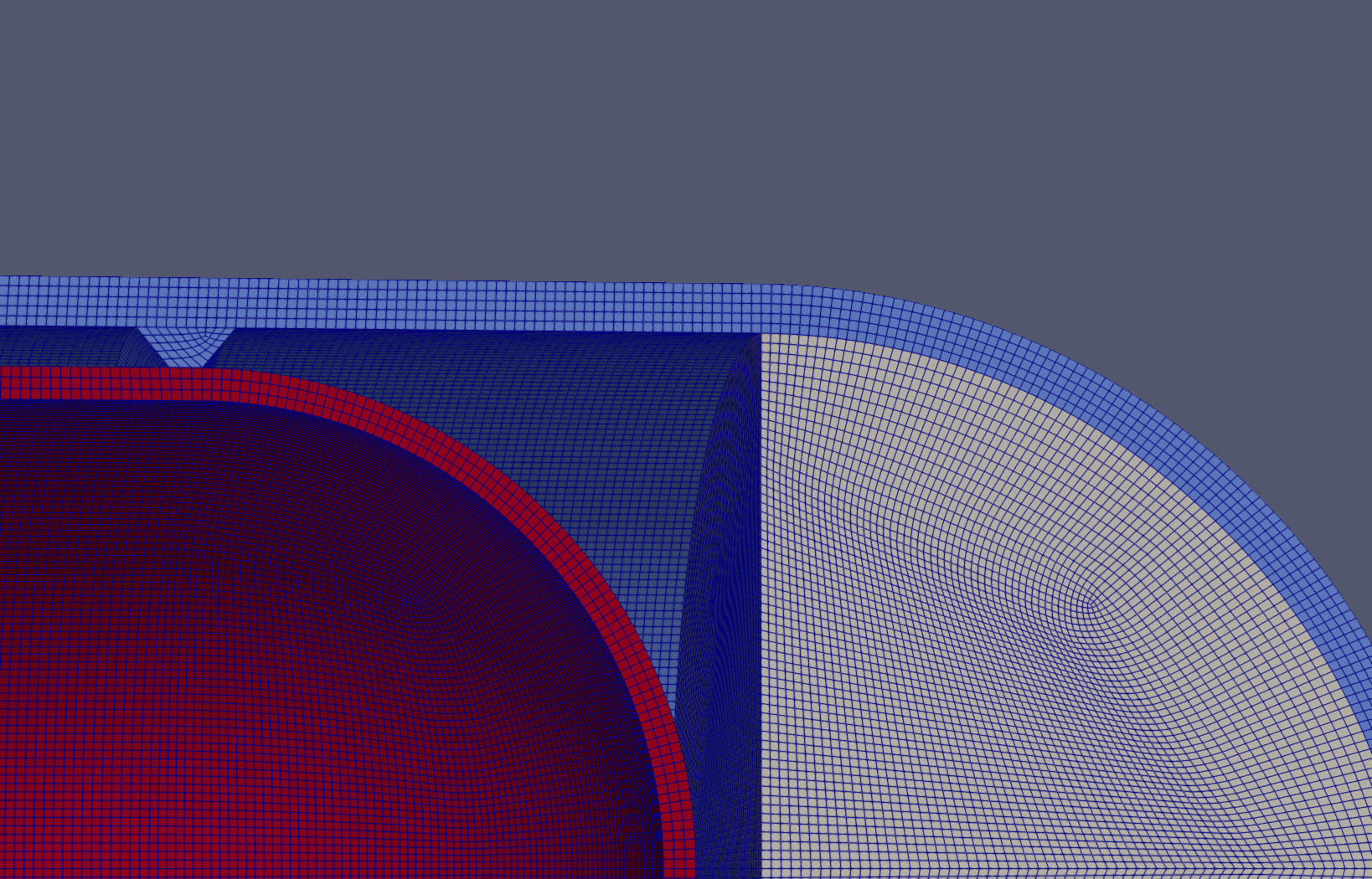}
    \includegraphics[width=0.4\textwidth]{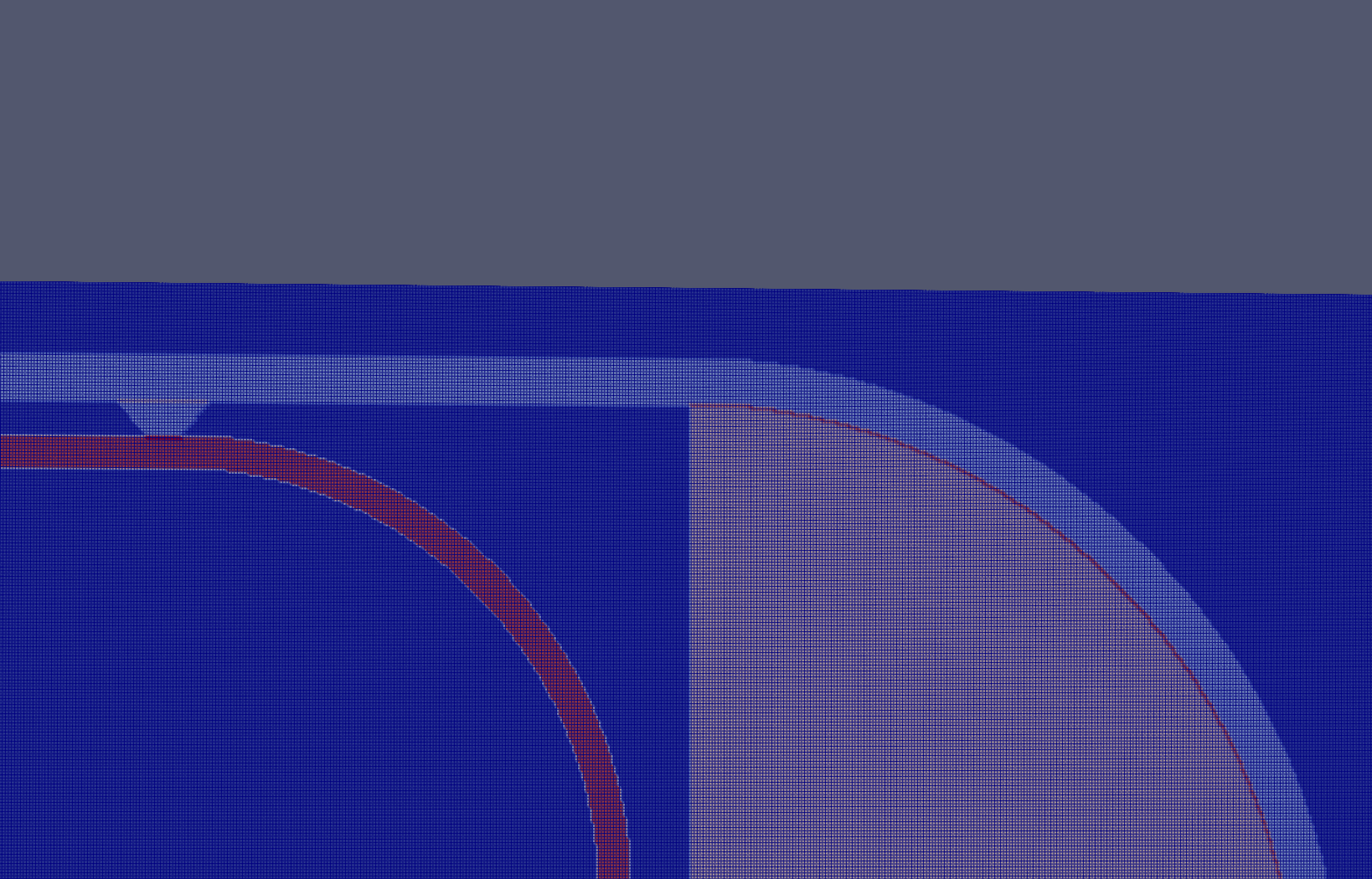}
    \caption{A snapshot of the conformal boundary mesh representing the rocket (left).  The resulting structured mesh after the first remap in the rocket problem run (right).}
\end{figure}

In the following benchmarks, the hardware we utilized was the Intel(R) Xeon(R) CPU E5-2695 v4 @ 2.10GHz.  Portage and its dependencies (Tangram etc.) were compiled with gcc/9.3.0.  The buffer sizes for J3D and R3D required by Portage and Tangram were used, that is, \texttt{J3D\_MAX\_VERTS=256}, \texttt{J3D\_MAX\_EDGES=512}, and \texttt{R3D\_MAX\_VERTS=1024}.  We retrieved the results from HPCToolkit \cite{HPCTK}.

\pgfplotstableread[row sep=\\,col sep=&]{
 method          & J3D & R3D \\
 total runtime   & 869 & 1720 \\
    }\mydata

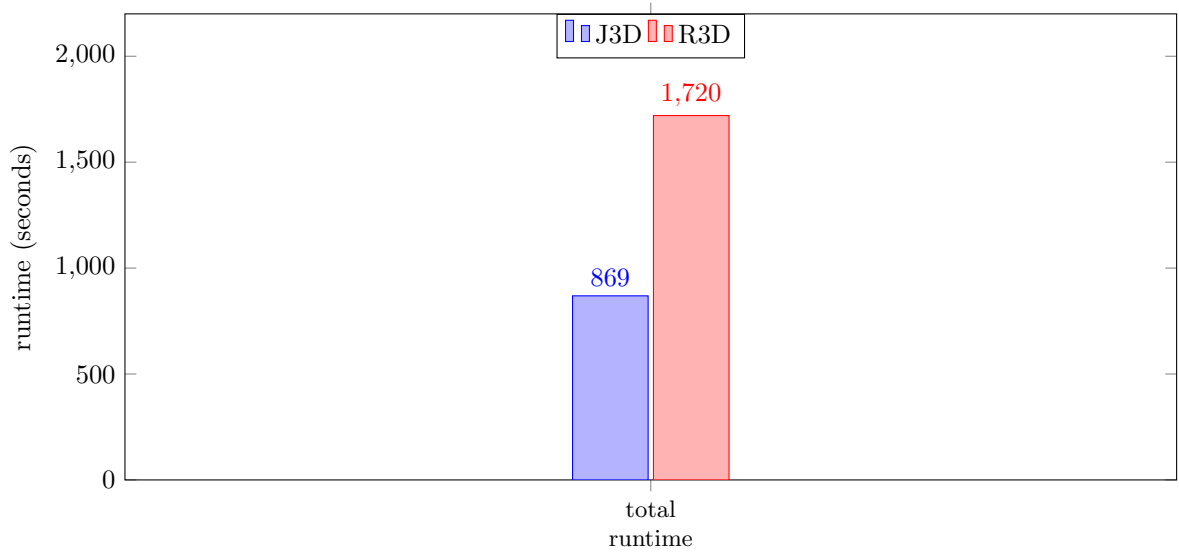
\begin{figure}[H]
\begin{tikzpicture}
    \begin{axis}[
            ybar,
            bar width=1cm,
            width=\textwidth,
            height=.5\textwidth,
            legend style={at={(0.5,1)},
                anchor=north,legend columns=-1},
            symbolic x coords={total runtime},
            x tick label style={font=\small,text width=1.7cm,align=center},
            xtick=data,
            nodes near coords,
            nodes near coords align={vertical},
            ymin=0,ymax=2200,
            ylabel={runtime (seconds)},
        ]
        \addplot table[x=method,y=J3D]{\mydata};
        \addplot table[x=method,y=R3D]{\mydata};
        \legend{J3D, R3D}
    \end{axis}
\end{tikzpicture}
\caption{Total runtime of the rocket problem with J3D vs R3D.} 
\end{figure}

\pgfplotstableread[row sep=\\,col sep=&]{
 method                     & J3D & R3D \\
 initialization             & 7.34 & 112 \\
 clip-n-cap                       & 253 & 366 \\
 moments computation        & 50.4 & 129 \\
    }\mydata

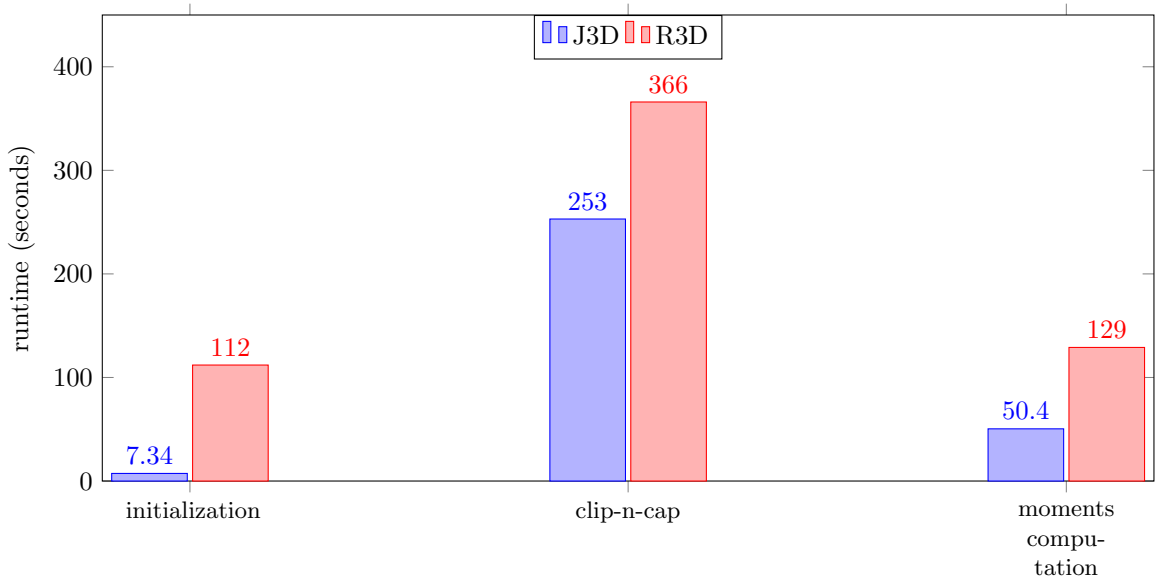
\begin{figure}[H]
\begin{tikzpicture}
    \begin{axis}[
            ybar,
            bar width=1cm,
            width=\textwidth,
            height=.5\textwidth,
            legend style={at={(0.5,1)},
                anchor=north,legend columns=-1},
            symbolic x coords={initialization,
                               clip-n-cap,
                               moments computation},
            x tick label style={font=\small,text width=1.7cm,align=center},
            xtick=data,
            nodes near coords,
            nodes near coords align={vertical},
            ymin=0,ymax=450,
            ylabel={runtime (seconds)},
        ]
        \addplot table[x=method,y=J3D]{\mydata};
        \addplot table[x=method,y=R3D]{\mydata};
        \legend{J3D, R3D}
    \end{axis}
\end{tikzpicture}
\caption{Runtime in seconds for each method in J3D and R3D after running the rocket problem.} 
\end{figure}

\section{Conclusion}
In this paper, we began by discussing the motivations for performant polyhedral clipping especially within the context of mesh to mesh interpolation and material interface reconstruction.  We then went into the implementation details of R3D as well as its performance limitations.  To address this, we introduced a novel use of data structures to implement more performant algorithms for clipping and we discussed some lessons learned in pursuit of this.  Finally, we presented some benchmarks on various problems which demonstrate J3D's impressive performance gains over R3D.

\section{Acknowledgments}
This work was conducted under the auspices of U.S. Government contract 89233218CNA000001 for Los Alamos National Laboratory, which is operated by Triad National Security, LLC for the U.S. Department of Energy/National Nuclear Security Administration.  We gratefully acknowledge the support of the U.S. Department of Energy National Nuclear Security Administration Advanced Simulation and Computing Program.  We would like to thank Jon Woodring for some helpful suggestions on the implementation of J3D.  We would like to thank Jacob Jones who contributed to the early discussions of efficient polyhedral intersection algorithms.  We would like to thank Evgeny Kikinzon who contributed the rotor problem benchmark implementation.  

\bibliographystyle{plain}
\bibliography{bib}

@article{LOUBERE20104724,
title = {ReALE: A reconnection-based arbitrary-Lagrangian–Eulerian method},
journal = {Journal of Computational Physics},
volume = {229},
number = {12},
pages = {4724-4761},
year = {2010},
issn = {0021-9991},
doi = {https://doi.org/10.1016/j.jcp.2010.03.011},
url = {https://www.sciencedirect.com/science/article/pii/S002199911000121X},
author = {Raphaël Loubère and Pierre-Henri Maire and Mikhail Shashkov and Jérôme Breil and Stéphane Galera}
}

@article{BARLOW2016603,
title = {Arbitrary Lagrangian–Eulerian methods for modeling high-speed compressible multimaterial flows},
journal = {Journal of Computational Physics},
volume = {322},
pages = {603-665},
year = {2016},
issn = {0021-9991},
doi = {https://doi.org/10.1016/j.jcp.2016.07.001},
url = {https://www.sciencedirect.com/science/article/pii/S0021999116302807},
author = {Andrew J. Barlow and Pierre-Henri Maire and William J. Rider and Robert N. Rieben and Mikhail J. Shashkov}
}

@article{powell2015r3d,
  title={r3d: Software for fast, robust geometric operations in 3D and 2D},
  author={Powell, D},
  journal={Los Alamos National Laboratory},
  year={2015}
}

@article{POWELL2015340,
title = {An exact general remeshing scheme applied to physically conservative voxelization},
journal = {Journal of Computational Physics},
volume = {297},
pages = {340-356},
year = {2015},
issn = {0021-9991},
doi = {https://doi.org/10.1016/j.jcp.2015.05.022},
url = {https://www.sciencedirect.com/science/article/pii/S0021999115003563},
author = {Devon Powell and Tom Abel}
}

@article{AHN20072096,
title = {Multi-material interface reconstruction on generalized polyhedral meshes},
journal = {Journal of Computational Physics},
volume = {226},
number = {2},
pages = {2096-2132},
year = {2007},
issn = {0021-9991},
doi = {https://doi.org/10.1016/j.jcp.2007.06.033},
url = {https://www.sciencedirect.com/science/article/pii/S0021999107002926},
author = {Hyung Taek Ahn and Mikhail Shashkov}
}

@ARTICLE{6127882,
  author={Koehl, Patrice},
  journal={IEEE Transactions on Pattern Analysis and Machine Intelligence}, 
  title={Fast Recursive Computation of 3D Geometric Moments from Surface Meshes}, 
  year={2012},
  volume={34},
  number={11},
  pages={2158-2163},
  doi={10.1109/TPAMI.2012.23}}

@article{Sugihara,
author = {Sugihara, Kokichi},
title = {A Robust and Consistent Algorithm for Intersecting Convex Polyhedra},
journal = {Computer Graphics Forum},
volume = {13},
number = {3},
pages = {45-54},
doi = {https://doi.org/10.1111/1467-8659.1330045},
url = {https://onlinelibrary.wiley.com/doi/abs/10.1111/1467-8659.1330045},
eprint = {https://onlinelibrary.wiley.com/doi/pdf/10.1111/1467-8659.1330045},
year = {1994}
}

@article{Herring-2021,
 author = {Herring, Angela and Ferenbaugh, Charles and Malone, Christopher and Shevitz, Daniel and Kikinzon, Evgeny and Dilts, Gary and Rakotoarivelo, Hoby and Velechovsky, Jan and Lipnikov, Konstantin and Ray, Navamita and Garimella, Rao},
 doi = {10.5334/jors.338},
 journal = {Journal of Open Research Software},
 month = {Sep},
 title = {Portage: A Modular Data Remap Library for Multiphysics Applications on Advanced Architectures},
 year = {2021}
}

@article{HPCTK,
author = {Adhianto, L. and Banerjee, S. and Fagan, M. and Krentel, M. and Marin, G. and Mellor-Crummey, J. and Tallent, N. R.},
title = {HPCTOOLKIT: tools for performance analysis of optimized parallel programs},
journal = {Concurrency and Computation: Practice and Experience},
volume = {22},
number = {6},
pages = {685-701},
doi = {https://doi.org/10.1002/cpe.1553},
url = {https://onlinelibrary.wiley.com/doi/abs/10.1002/cpe.1553},
eprint = {https://onlinelibrary.wiley.com/doi/pdf/10.1002/cpe.1553},
year = {2010}
}

@article{dyadechko2005moment,
  title={Moment-of-fluid interface reconstruction},
  author={Dyadechko, Vadim and Shashkov, Mikhail},
  journal={Los Alamos Report LA-UR-05-7571},
  volume={49},
  year={2005}
}

@article{VOF,
    author = {Benson , David J },
    title = {Volume of fluid interface reconstruction methods for multi-material problems},
    journal = {Applied Mechanics Reviews},
    volume = {55},
    number = {2},
    pages = {151-165},
    year = {2002},
    month = {04},
    issn = {0003-6900},
    doi = {10.1115/1.1448524},
    url = {https://doi.org/10.1115/1.1448524},
    eprint = {https://asmedigitalcollection.asme.org/appliedmechanicsreviews/article-pdf/55/2/151/5439062/151\_1.pdf},
}

@conference{lviraplus,
title = {LVIRA+, extended least squares Volume-of-Fluid interface reconstruction algorithm for m $>$ 2 material domains},
author = {Evgeny Kikinzon and Hoby Nirina Rakotoarivelo and Jan Velechovsky},
number = {LA-UR-24-29017},
note = {Sponsor: USDOE Office of Science (SC). Advanced Scientific Computing Research (ASCR) (SC-31)},
year = {2024},
note = {International Conference on Numerical Methods for Multi-Material Fluid Flow (MultiMat) ; 2024-08-26 - 2024-08-30 ; Breckenridge, Colorado, United States},
howpublished = {International Conference on Numerical Methods for Multi-Material Fluid Flow (MultiMat) ; 2024-08-26 - 2024-08-30 ; Breckenridge, Colorado, United States},
note = {Version: 1},
url = {https://laauthors.lanl.gov/record/file?id=perma%3Alanl%2Frl-dv-0000477009%2F2},
note = {Repository ID: perma:lanl/rl-dv-0000477009},
note = {Dataset: rassti},
}

@article{LVIRA,
title = {Second-order accurate volume-of-fluid algorithms for tracking material interfaces},
journal = {Journal of Computational Physics},
volume = {199},
number = {2},
pages = {465-502},
year = {2004},
issn = {0021-9991},
doi = {https://doi.org/10.1016/j.jcp.2003.12.023},
url = {https://www.sciencedirect.com/science/article/pii/S0021999104000920},
author = {James Edward Pilliod and Elbridge Gerry Puckett}
}

@article{e756a2bc-5e46-3b95-8a3a-c165473da678,
 ISSN = {00255718, 10886842},
 URL = {http://www.jstor.org/stable/2003308},
 author = {Gilbert C. Best},
 journal = {Mathematics of Computation},
 number = {86},
 pages = {310--312},
 publisher = {American Mathematical Society},
 title = {Helpful Formulas for Integrating Polynomials in Three Dimensions},
 urldate = {2025-08-13},
 volume = {18},
 year = {1964}
}

@inproceedings{CSR,
author = {Bulu\c{c}, Aydin and Fineman, Jeremy T. and Frigo, Matteo and Gilbert, John R. and Leiserson, Charles E.},
title = {Parallel sparse matrix-vector and matrix-transpose-vector multiplication using compressed sparse blocks},
year = {2009},
isbn = {9781605586069},
publisher = {Association for Computing Machinery},
address = {New York, NY, USA},
url = {https://doi.org/10.1145/1583991.1584053},
doi = {10.1145/1583991.1584053},
booktitle = {Proceedings of the Twenty-First Annual Symposium on Parallelism in Algorithms and Architectures},
pages = {233–244},
numpages = {12},
location = {Calgary, AB, Canada},
series = {SPAA '09}
}

@article{MULLER1978217,
title = {Finding the intersection of two convex polyhedra},
journal = {Theoretical Computer Science},
volume = {7},
number = {2},
pages = {217-236},
year = {1978},
issn = {0304-3975},
doi = {https://doi.org/10.1016/0304-3975(78)90051-8},
url = {https://www.sciencedirect.com/science/article/pii/0304397578900518},
author = {D.E. Muller and F.P. Preparata}
}

@article{Chazelle,
author = {Chazelle, Bernard},
title = {An Optimal Algorithm for Intersecting Three-Dimensional Convex Polyhedra},
journal = {SIAM Journal on Computing},
volume = {21},
number = {4},
pages = {671-696},
year = {1992},
doi = {10.1137/0221041},

URL = { https://doi.org/10.1137/0221041},
eprint = { https://doi.org/10.1137/0221041}
}

@article{Bernstein,
journal = {Computer Graphics Forum},
title = {{Fast, Exact, Linear Booleans}},
author = {Bernstein, Gilbert and Fussell, Don},
year = {2009},
publisher = {The Eurographics Association and Blackwell Publishing Ltd},
ISSN = {1467-8659},
DOI = {10.1111/j.1467-8659.2009.01504.x}
}

\appendix

\section{Formula for Explicit Moments Computation}
To properly define the moments, we will begin by considering three points from a triangular facet $A$, $B$, and $C$, along with the origin $O$, which form a tetrahedron $\sigma_{T}=(O,A,B,C)$.  The geometric moments of the tetrahedron are the following, from Koehl \cite{6127882}.
\begin{equation}
M_{i,j,k}=sign(vol(\sigma_{T}))\int_{\sigma_{T}}x^{i}y^{j}z^{k}dxdydz
\end{equation}
The integral over the tetrahedron $\sigma_{T}$ is the following \cite{e756a2bc-5e46-3b95-8a3a-c165473da678}:
\begin{equation}
\int_{\sigma_{T}}x^{i}y^{j}z^{k}dxdydz=\frac{6|vol(\sigma)|}{(i+j+k+3)!}S_{i,j,k}(\sigma_{T})
\end{equation}
where
\begin{equation}
\begin{matrix}
S_{i,j,k}(\sigma_{T})=\sum_{(i_{1}+i_{2}+i_{3}=i)}\sum_{(j_{1}+j_{2}+j_{3}=j)}\sum_{(k_{1}+k_{2}+k_{3}=k)} \\
\frac{(i_{1}+j_{1}+k_{1})!}{i_{1}!j_{1}!k_{1}!}x_{A}^{i_{1}}y_{A}^{j_{1}}z_{A}^{k_{1}} \\
\frac{(i_{2}+j_{2}+k_{2})!}{i_{2}!j_{2}!k_{2}!}x_{B}^{i_{2}}y_{B}^{j_{2}}z_{B}^{k_{2}} \\
\frac{(i_{3}+j_{3}+k_{3})!}{i_{3}!j_{3}!k_{3}!}x_{C}^{i_{3}}y_{C}^{j_{3}}z_{C}^{k_{3}}
\end{matrix}
\end{equation}
and the oriented volume, $vol(\sigma_{T})$, of the tetrahedron $\sigma_{T}$ is given by
\begin{equation}
vol(\sigma_{T})=\frac{1}{6}
\begin{vmatrix}
x_{A} & y_{A} & z_{A}\\
x_{B} & y_{B} & z_{B}\\
x_{C} & y_{C} & z_{C}
\end{vmatrix}.
\end{equation}
By combining equations (2) and (3) we have that the geometric moments of the tetrahedron are
\begin{equation}
M_{i,j,k}=6\frac{i!j!k!}{(i+j+k+3)!}vol(\sigma_{T})S_{i,j,k}(\sigma_{T}).
\end{equation}
Using equation (6) the explicit moments of a tetrahedra from zero through third order are the following
\begin{equation}
M_{0,0,0}=vol(\sigma_{T})
\end{equation}
\begin{equation}
M_{1,0,0}=\frac{1}{4}vol(\sigma_{T})(x_{A}+x_{B}+x_{C})
\end{equation}
\begin{equation}
M_{0,1,0}=\frac{1}{4}vol(\sigma_{T})(y_{A}+y_{B}+y_{C})
\end{equation}
\begin{equation}
M_{0,0,1}=\frac{1}{4}vol(\sigma_{T})(z_{A}+z_{B}+z_{C})
\end{equation}
\begin{equation}
M_{2,0,0}=\frac{1}{10}vol(\sigma_{T})(x_A^2 + x_A x_B + x_B^2 + x_A x_C + x_B x_C + x_C^2)
\end{equation}
\begin{equation}
M_{1,1,0}=\frac{1}{20}vol(\sigma_{T})(x_A^2 + x_A x_B + x_B^2 + x_A x_C + x_B x_C + x_C^2)
\end{equation}
\begin{equation}
M_{1,0,1}=\frac{1}{20}vol(\sigma_{T})(2 x_A z_A + z_A x_B + x_A z_B + 2 x_B z_B + z_A x_C + z_B x_C + 
 x_A z_C + x_B z_C + 2 x_C z_C)
\end{equation}
\begin{equation}
M_{0,2,0}=\frac{1}{10}vol(\sigma_{T})(y_A^2 + y_A y_B + y_B^2 + y_A y_C + y_B y_C + y_C^2)
\end{equation}
\begin{equation}
M_{0,1,1}=\frac{1}{20}vol(\sigma_{T})(2 y_A z_A + z_A y_B + y_A z_B + 2 y_B z_B + z_A y_C + z_B y_C + 
 y_A z_C + y_B z_C + 2 y_C z_C)
\end{equation}
\begin{equation}
M_{0,0,2}=\frac{1}{10}vol(\sigma_{T})(z_A^2 + z_A z_B + z_B^2 + z_A z_C + z_B z_C + z_C^2)
\end{equation}
\begin{equation}
M_{3,0,0}=\frac{1}{20}vol(\sigma_{T})(x_A^3 + x_A^2 x_B + x_A x_B^2 + x_B^3 + x_A^2 x_C + x_A x_B x_C + 
 x_B^2 x_C + x_A x_C^2 + x_B x_C^2 + x_C^3)
\end{equation}
\begin{equation}
\begin{matrix}
M_{2,1,0}=\frac{1}{60}vol(\sigma_{T})(3 x_A^2 y_A + 2 x_A y_A x_B + y_A x_B^2 + x_A^2 y_B + 2 x_A x_B y_B + 3 x_B^2 y_B \\
+ 2 x_A y_A x_C + y_A x_B x_C + x_A y_B x_C + 2 x_B y_B x_C + y_A x_C^2 + y_B x_C^2 \\
+ x_A^2 y_C + x_A x_B y_C + x_B^2 y_C + 2 x_A x_C y_C + 2 x_B x_C y_C + 3 x_C^2 y_C)
\end{matrix}
\end{equation}
\begin{equation}
\begin{matrix}
M_{2,0,1}=\frac{1}{60}vol(\sigma_{T})(3 x_A^2 z_A + 2 x_A z_A x_B + z_A x_B^2 + x_A^2 z_B + 2 x_A x_B z_B + 3 x_B^2 z_B \\
+ 2 x_A z_A x_C + z_A x_B x_C + x_A z_B x_C + 2 x_B z_B x_C + z_A x_C^2 + z_B x_C^2 \\
+ x_A^2 z_C + x_A x_B z_C + x_B^2 z_C + 2 x_A x_C z_C + 2 x_B x_C z_C + 3 x_C^2 z_C)
\end{matrix}
\end{equation}
\begin{equation}
\begin{matrix}
M_{1,2,0}=\frac{1}{60}vol(\sigma_{T})(3 x_A y_A^2 + y_A^2 x_B + 2 x_A y_A y_B + 2 y_A x_B y_B + x_A y_B^2 + 3 x_B y_B^2 \\
+ y_A^2 x_C + y_A y_B x_C + y_B^2 x_C + 2 x_A y_A y_C + y_A x_B y_C + x_A y_B y_C \\
+ 2 x_B y_B y_C + 2 y_A x_C y_C + 2 y_B x_C y_C + x_A y_C^2 + x_B y_C^2 + 3 x_C y_C^2)
\end{matrix}
\end{equation}
\begin{equation}
\begin{matrix}
M_{1,1,1}=\frac{1}{120}vol(\sigma_{T})(6 x_A y_A z_A + 2 y_A z_A x_B + 2 x_A z_A y_B + 2 z_A x_B y_B + 2 x_A y_A z_B + 2 y_A x_B z_B \\
+ 2 x_A y_B z_B + 6 x_B y_B z_B + 2 y_A z_A x_C + z_A y_B x_C + y_A z_B x_C + 2 y_B z_B x_C \\
+ 2 x_A z_A y_C + z_A x_B y_C + x_A z_B y_C + 2 x_B z_B y_C + 2 z_A x_C y_C + 2 z_B x_C y_C \\
+ 2 x_A y_A z_C + y_A x_B z_C + x_A y_B z_C + 2 x_B y_B z_C + 2 y_A x_C z_C + 2 y_B x_C z_C \\
+ 2 x_A y_C z_C + 2 x_B y_C z_C + 6 x_C y_C z_C)
\end{matrix}
\end{equation}
\begin{equation}
\begin{matrix}
M_{1,0,2}=\frac{1}{60}vol(\sigma_{T})(3 x_A z_A^2 + z_A^2 x_B + 2 x_A z_A z_B + 2 z_A x_B z_B + x_A z_B^2 + 3 x_B z_B^2 \\
+ z_A^2 x_C + z_A z_B x_C + z_B^2 x_C + 2 x_A z_A z_C + z_A x_B z_C + x_A z_B z_C \\
+ 2 x_B z_B z_C + 2 z_A x_C z_C + 2 z_B x_C z_C + x_A z_C^2 + x_B z_C^2 + 3 x_C z_C^2)
\end{matrix}
\end{equation}
\begin{equation}
M_{0,3,0}=\frac{1}{20}vol(\sigma_{T})(y_A^3 + y_A^2 y_B + y_A y_B^2 + y_B^3 + y_A^2 y_C + y_A y_B y_C + 
 y_B^2 y_C + y_A y_C^2 + y_B y_C^2 + y_C^3)
\end{equation}
\begin{equation}
\begin{matrix}
M_{0,2,1}=\frac{1}{60}vol(\sigma_{T})(3 y_A^2 z_A + 2 y_A z_A y_B + z_A y_B^2 + y_A^2 z_B + 2 y_A y_B z_B + 3 y_B^2 z_B \\
+ 2 y_A z_A y_C + z_A y_B y_C + y_A z_B y_C + 2 y_B z_B y_C + z_A y_C^2 + z_B y_C^2 \\
+ y_A^2 z_C + y_A y_B z_C + y_B^2 z_C + 2 y_A y_C z_C + 2 y_B y_C z_C + 3 y_C^2 z_C)
\end{matrix}
\end{equation}
\begin{equation}
\begin{matrix}
M_{0,1,2}=\frac{1}{60}vol(\sigma_{T})(3 y_A z_A^2 + z_A^2 y_B + 2 y_A z_A z_B + 2 z_A y_B z_B + y_A z_B^2 + 3 y_B z_B^2 \\
+ z_A^2 y_C + z_A z_B y_C + z_B^2 y_C + 2 y_A z_A z_C + z_A y_B z_C + y_A z_B z_C \\
+ 2 y_B z_B z_C + 2 z_A y_C z_C + 2 z_B y_C z_C + y_A z_C^2 + y_B z_C^2 + 3 y_C z_C^2)
\end{matrix}
\end{equation}
\begin{equation}
M_{0,0,3}=\frac{1}{20}vol(\sigma_{T})(z_A^3 + z_A^2 z_B + z_A z_B^2 + z_B^3 + z_A^2 z_C + z_A z_B z_C + 
 z_B^2 z_C + z_A z_C^2 + z_B z_C^2 + z_C^3)
\end{equation}

\end{document}